%% file: network.tex
\documentclass[11 pt, onecolumn]{article}
\usepackage[margin=1.1in]{geometry}

\input{./sections/initialize}
\usepackage{algorithm}
\usepackage{algpseudocode}
\usepackage{caption}
\usepackage{mathrsfs}

\DeclareMathAlphabet{\mathcal}{OMS}{cmsy}{m}{n}

\usepackage{graphicx}
\usepackage{epstopdf}

\begin{document}
\title{\huge \bf An Approximate Dynamic Programming Approach to Vehicle Platooning Coordination in Networks}

\author{Xi Xiong \thanks{Key Laboratory of Road and Traffic Engineering, Ministry of Education, Tongji University, Shanghai, China}   ,
Maonan Wang \thanks{School of Science and Engineering, The Chinese University of Hong Kong, Shenzhen, China}   ,
Dengfeng Sun \thanks{School of Aeronautics and Astronautics, Purdue University, USA.}   ,
Li Jin \thanks{UM Joint Institute and the Department of Automation, Shanghai Jiao Tong University, China.},
\footnote{Corresponding author: li.jin@sjtu.edu.cn.}
}



\newcommand*{\QEDA}{\hfill\ensuremath{\blacksquare}}%

\maketitle

\begin{abstract}
Platooning connected and autonomous vehicles (CAVs) provide significant benefits in terms of traffic efficiency and fuel economy.
However, most existing platooning systems assume the availability of pre-determined plans, which is not feasible in real-time scenarios.
In this paper, we address this issue in time-dependent networks by formulating a Markov decision process at each junction, aiming to minimize travel time and fuel consumption.
Initially, we analyze coordinated platooning without routing to explore the cooperation among controllers on an identical path. 
We propose two novel approaches based on approximate dynamic programming, offering suboptimal control in the context of a stochastic finite horizon problem.
The results demonstrate the superiority of the approximation in the policy space.
Furthermore, we investigate platooning in a network setting, where speed profiles and routes are determined simultaneously.
To simplify the problem, we decouple the action space by prioritizing routing decisions based on travel time estimation.
We subsequently employ the aforementioned policy approximation to determine speed profiles, considering essential parameters such as travel times.
Our simulation results in SUMO indicate that our method yields better performance than conventional approaches, leading to potential travel cost savings of up to $40\%$.
Additionally, we evaluate the resilience of our approach in dynamically changing networks, affirming its ability to maintain efficient platooning operations.
\end{abstract}

{\bf Index terms}: Vehicle platooning, Multi-agent systems, Approximate dynamic programming, Bellman equation.

\input{./sections/01_introduction} 
\input{./sections/02_modeling} 
\input{./sections/03_cascade} 
\input{./sections/04_network} 
\input{./sections/05_conclusion} 

\newpage

\section*{Acknowledgments}
This work was supported in part by C2SMART University Transportation Center, NYU Tandon School of Engineering, NSFC Project 62103260, SJTU UM Joint Institute, J. Wu \& J. Sun Endowment Fund, and Fundamental Research Funds for the Central Universities.

\bibliographystyle{IEEEtran}
\bibliography{network}

\end{document}

%% file: sections/initialize.tex
\usepackage{subfigure}
\usepackage{amsmath,amssymb}

\usepackage{amsthm} 
\allowdisplaybreaks
\usepackage{color}
\usepackage{footnote}
\usepackage{algorithm}
\usepackage{algpseudocode}
\usepackage{algorithmicx}
\usepackage{multirow} 
\usepackage{booktabs}
\usepackage{graphicx}
\usepackage{amssymb}
\usepackage{amsbsy}
\usepackage{array}
\usepackage{longtable}
\usepackage{epstopdf}
\usepackage{pbox}
\usepackage{url}
\usepackage{breqn}
\usepackage{mathrsfs}
\usepackage{multicol}
\usepackage{supertabular}
\usepackage{enumerate}
\usepackage{bbm}
\usepackage{multirow}
\usepackage{hyperref}
\usepackage{cite}
\usepackage{xfrac}
\usepackage{textgreek}
\usepackage[justification=centering]{caption}
\usepackage{mathtools}

\usepackage{todonotes}
\usepackage{dsfont}
\newcommand{\argmin}{\arg\!\min}

\algnewcommand\algorithmicforeach{\textbf{for each}}
\algdef{S}[FOR]{ForEach}[1]{\algorithmicforeach\ #1\ \algorithmicdo}

%% file: sections/01_introduction.tex
\section{Introduction}

Connected and autonomous vehicles (CAVs) are rapidly advancing due to recent progress in artificial intelligence and wireless communication. Platooning enables vehicles to travel together with very small inter-vehicle headways, resulting in improved traffic efficiency and fuel economy, particularly for heavy-duty vehicles \cite{axelsson2016safety, tsugawa2016review}. A typical platoon management system comprises three hierarchies: the vehicle layer, junction layer, and network layer. While previous efforts have primarily concentrated on the vehicle layer, focusing on  microscopic longitudinal and lateral control schemes \cite{kong2017millimeter, zhou2019robust, turri2016cooperative, talebpour2016influence}, recent research has explored coordinated platooning at the junction and network layers to minimize travel costs \cite{rios2016survey, bhoopalam2018planning}. However, a common limitation of this research is the assumption that itinerary plans are static and not responsive to real-time conditions.

\begin{figure}[htb]
  \centering
  {
  \includegraphics[width=0.56\columnwidth,trim=650 320 400 70,clip]{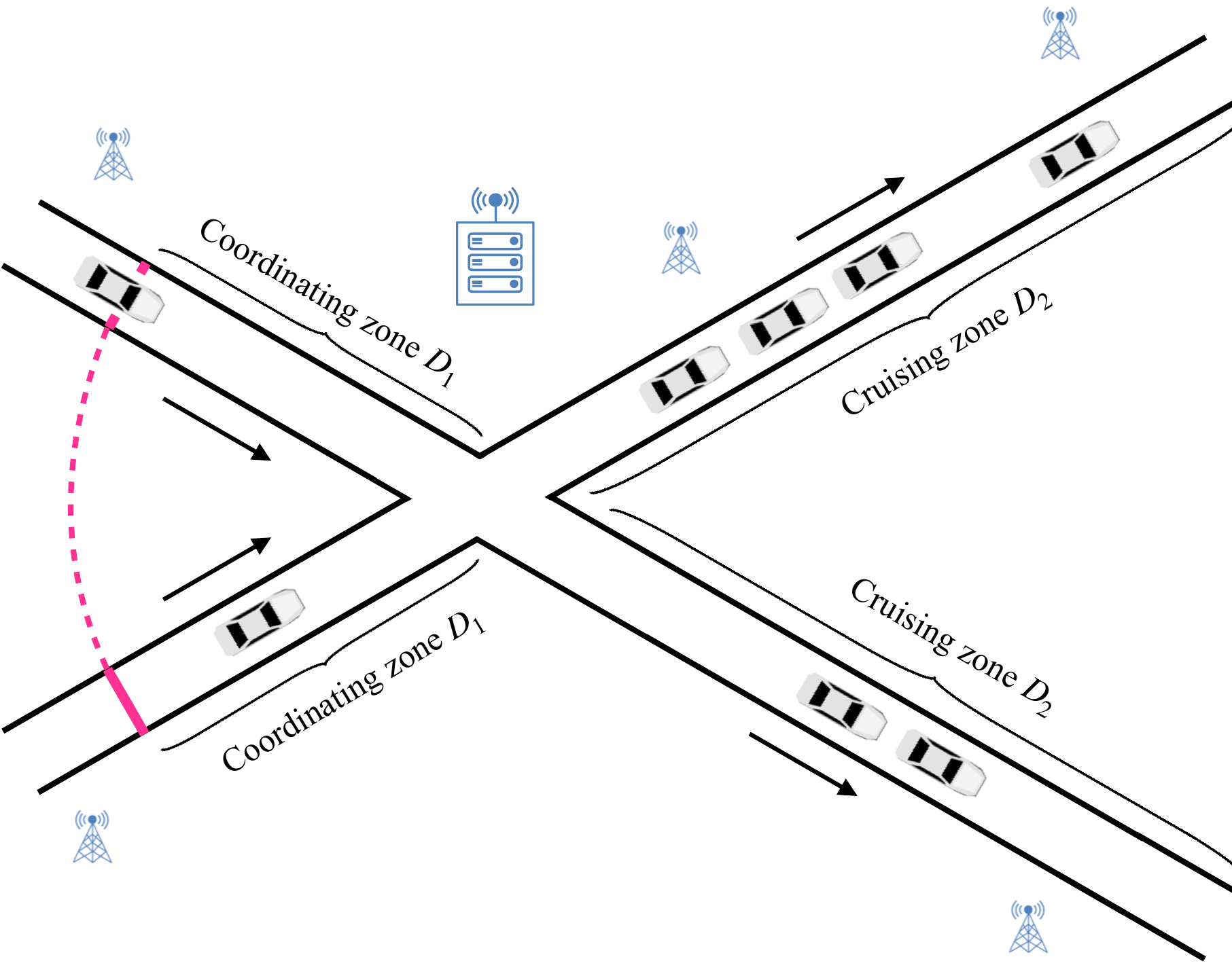}
  }
  \caption{Platooning coordination at a highway junction.} \label{fig:color_figure}
\end{figure}

To enhance the platoon management system, it is crucial to integrate real-time responsiveness into the itinerary planning process, which holds the potential for improving resource utilization and overall system performance.
In this paper, we mainly focus on scalable platooning coordination in time-dependent networks, employing an approximate dynamic programming approach.
Fig.~\ref{fig:color_figure} illustrates the platooning coordination at a highway junction. The solid red lines represent detectors that span the road segment, enabling the collection of trip information such as destinations, locations and arrival times. These data can be shared through a communication infrastructure such as a 5G cellular network. The edges in the diagram are divided into two distinct sections: the \textit{coordinating zone} and the \textit{cruising zone}. Within the coordinating zone, detectors transmit the trip information to the virtual controller located at the junction. The primary role of the virtual controller is to determine whether the current CAV should merge with the CAV ahead to form a platoon.
Upon passing the junction, the newly formed platoon proceeds to travel together in the cruising zone, with the following vehicle
benefiting from fuel economy advantages resulting from reduced air resistance.
The platooning strategy is closely tied to path planning in networks, as we focus on the balance between travel time and fuel consumption.
To facilitate effective cooperation among controllers, we develop an approximate dynamic programming approach for platooning coordination in time-dependent networks. This approach enables effective coordination mechanisms and address the challenges associated with real-time strategies.




Previous research on platooning coordination in networks has predominantly focused on off-line planning, where all trips are announced before the planning phase. For small-scale cases, exact solutions have been proposed, while heuristic methods have been devised to tackle larger instances \cite{bhoopalam2018planning}.
Larson et al. \cite{larson2013coordinated} were the pioneers in studying scalable coordination by introducing distributed controllers throughout a road network. Their work was subsequently extended in \cite{larson2015distributed}. 
In another study, Larson et al. \cite{larsson2015vehicle} proved that the network-level platooning problem was NP-hard and proposed integer linear programming formulations for instances where deadlines are discarded.
Abdolmaleki et al. \cite{abdolmaleki2019itinerary} formulated the problem as a minimum concave-cost network flow problem and designed an outer approximation algorithm for solving a mixed-integer convex minimization reformulation.
Nourmohammadzadeh and Hartmann \cite{nourmohammadzadeh2016fuel} focused on the platooning problem with individual deadlines for vehicles and employed a Genetic Algorithm to find solutions for larger instances.
Van de Hoef et al. \cite{van2017fuel} proposed a combinatorial optimization approach that combines plans involving only two vehicle to solve the platooning problem and developed a heuristic algorithm for larger instances.
Luo and Larson \cite{luo2022repeated} proposed an iterative route-then-schedule heuristic for centralized planning, which exhibited rapid convergence to high-quality solutions.
Furthermore, researchers have addressed practical constraints such as communication systems, travel time uncertainties
behavioral-instability issues, and mandatory breaks for drivers \cite{balador2022survey, zhang2017freight, sun2021decentralized, xu2022truck}. However, these approaches assume a priori knowledge of all vehicles' trip schedules, disregarding the fact that platoons often split and recombine, leading to changes in the system's state \cite{bhoopalam2018planning}.
To the best of our knowledge, this paper represents the first effort to consider dynamic planning, where decisions need to be made as vehicles are en-route.

We formulate the problem over a multi-origin-destination (O-D) network.
When a CAV enters a junction's coordinating zone, the controller receives the trip information to determine speed profiles and optimal routes.
To tackle this junction-level coordination, we adopt a Markov decision process (MDP) framework. 
At each junction, the state comprises headways and destinations, while the action involves selecting the optimal path and assigning a speed.
Our primary objective is to minimize the system-level cost, which encompasses travel time and fuel consumption. In our formulation, we account for the presence of mixed-autonomy traffic, which aligns with the setting explored in \cite{segata2014supporting}. Furthermore, we incorporate a non-linear cost function to capture the impact of dynamic flows on travel cost, similar to the model presented in \cite{abdolmaleki2019itinerary}.

In this paper, we first analyze coordinated platooning without routing, aiming to explore the cooperation among controllers on an identical path.
It is difficult to find exact solutions resulting from the interdependencies among controllers' decisions.
To address this, we propose two general approaches based on dynamic programming for suboptimal control in the context of a stochastic finite horizon problem.
The first approach involves approximating cost functions by incorporating the cost estimations of adjacent vertices obtained from a dynamic programming equation. 
In the second approach, we constrain the policy to lie within a given parametric class, involving travel times and cruising distances. 
Additionally, we compare the performance of the two methods in the simulation platform, and the results demonstrate the superiority of the approximation in the policy space.

Based on the analysis presented above, we study real-time coordinated platooning in time-dependent networks, where the controller at each junction determines speed profiles and routes simultaneously. 
To simplify the problem structure, we decouple the hybrid action space, giving priority to the routing decision based on the observation that vehicles generally prefer to minimize their travel time \cite{larson2015distributed}.
The speed profiles for platooning are determined through the aforementioned approximation in the policy space, considering relevant parameters such as travel times and cruising distances.
Furthermore, we evaluate our approach using the Nguyen-Dupuis network in the simulation platform.
The results demonstrate that our method outperforms conventional methods, with remarkable reductions in travel costs of up to $40\%$ during heavy congestion periods.
Additionally, we assess the performance of our method in dynamically changing networks by simulating edge disconnections, and the results indicate its ability to effectively react to such changes and maintain efficient platooning operations. The main contributions of this paper are summarized as follows:
\begin{enumerate}
    \item We formulate a MDP framework for junction-level coordination aimed at minimizing travel costs consisting of travel time and fuel consumption.
    \item We study coordination without routing to analyze coordination among controllers and propose two general approaches based on dynamic programming for suboptimal control. The simulation results affirm the superiority of the approximation in the policy space.
    \item We decouple the action space for real-time coordination in networks. This approach prioritizes the routing decision with travel time estimation while employing the aforementioned policy approximation to determine speed profiles. The simulation results demonstrate the efficiency and resiliency of our approach.
\end{enumerate}

The remaining sections of this paper are organized as follows. 
In Section~\ref{sec_model}, we present the modeling of platooning coordination and formulate the Markov decision process to handle coordination at junctions.
In Section~\ref{sec_cascade}, we focus on the analysis of coordinated platooning without routing across a cascade of junctions.
In Section~\ref{sec_network}, we address the consideration of coordinated platooning and adaptive routing concurrently within networks.
In Section~\ref{sec_conclude}, we summarize the conclusions and propose several directions for future work.

%% file: sections/02_modeling.tex
\section{Modeling and Formulation}
\label{sec_model}

In this section, we first introduce the modeling of scalable platooning coordination and microscopic car-following models, and then present the problem definition.


\subsection{Modeling}
\subsubsection{Platooning coordination in networks}
\label{Section: preliminary definitions}
\begin{figure}[htbp]
  \centering
  \includegraphics[width=0.67\textwidth, trim=400 270 380 270,clip ]{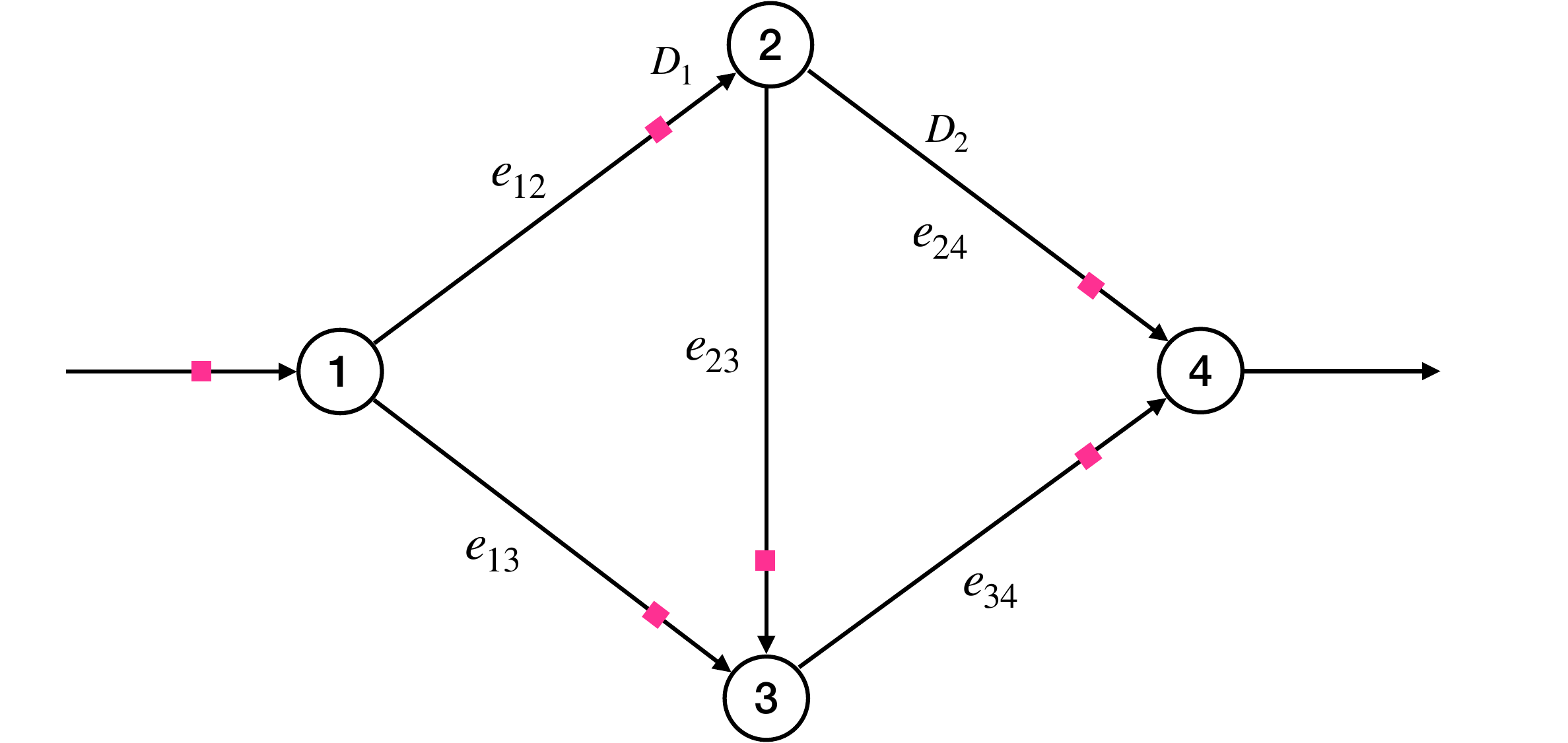}
  \caption{Platooning coordination in a simple network}
  \label{fig:small network}
\end{figure}

Consider a directed network $G = (V, E)$ with vertices $V$ and edges $E$. The number of vertices is represented as $N = | V |$, while the number of edges is denoted as $L = |E|$. We define a vertex $\nu \in V$ with more than two edges as a junction. Let $e_{ij}$ denote the road segment connecting vertex $\nu_i$ and vertex $\nu_j$. 
Initially, our focus lies on the platooning coordination at these junctions, where virtual controllers situated at the junctions determine the formation of platoons. Fig.~\ref{fig:small network} shows a simple network with platooning coordination. There is a coordinating zone $D_1$ before each junction. The red points denote the detectors at the start of each $D_1$, which can collect trip information such as destinations, locations and arrival times. After each junction, vehicles travel in platoons or alone in the cruising zone $D_2$. An edge is therefore divided into 2 segments: the coordinating zone $D_1$ and the cruising zone $D_2$. In practice, $D_1$ is much less than $D_2$ due to limited communication capabilities.

When vehicle $k$ enters the coordinating zone of vertex $\nu_i$, the time of entry $t^k_i$ and its destination $d^k$ are transmitted to the virtual controller at vertex $\nu_i$. The controller then determines whether vehicle $k$ merges with its leading vehicle $k-1$ based on kinematic information from both vehicles. We assume that the nominal speed in the network is $v_0$. When the controller instructs vehicle $k$ to merge with vehicle $k-1$, vehicle $k$ will adjust its speed to meet with vehicle $k-1$ at vertex $\nu_i$, then they will travel in groups in the following cruising zone. We make the practical assumption that vehicle $k$ will not directly influence the movement of vehicle $k-1$. Additionally, vehicle $k$ will travel in the coordinating zone with a reference speed from the controller when it is instructed to travel alone.

In network-level coordination, the controller at each junction determines speed profiles and routes simultaneously. When the controller at vertex $\nu_i$ instructs vehicle $k$ to merge with vehicle $k-1$, vehicle $k$ traverses vertex $\nu_i$'s coordinating zone with the constant speed $v_i^k$ from $t_i^k$, and then meets with vehicle $k-1$ at vertex $\nu_i$. After the junction, vehicle $k$ follows vehicle $k-1$ on edge $e_{ij}$, which was determined when vehicle $k-1$ entered vertex $\nu_i$'s coordinating zone, until the newly formed platoon enters the coordinating zone of vertex $\nu_j$ at $t_j^k$.
We assume that all information can be shared through the communication infrastructure along the highway system. A network-level coordination framework is introduced in Section~\ref{sec_network} to facilitate cooperation among controllers.

\subsubsection{Microscopic car-following models}
We consider the microscopic car-following model Intelligent Driver Model (IDM) developed by Treiber \cite{treiber2017intelligent}. The vehicle dynamics of positions and velocities in IDM are shown as
\begin{equation*}
\label{Eq: vehicle_dynamics}
\begin{aligned}
    & \dot{q}_{k} = \frac{\mathrm{d}q_k}{\mathrm{d}t}, \\
    & \dot{v}_{k} = \frac{\mathrm{d}v_k}{\mathrm{d}t} = a \left(1 - \left(\frac{v_k}{v_0} \right)^\delta - \left(\frac{s^*\left(v_k, \Delta v_k \right)}{s_k} \right)^2 \right),
\end{aligned}
\end{equation*}
where $q_k$ is the position of vehicle $k$ at time step $t$, and $v_k$ is the vehicle speed. $s^*\left(v_k, \Delta v_k \right) = s_0 + v_k T + \frac{v_k \Delta v_k}{2 \sqrt{ab}}$, and $v_0$, $s_0$, $T$, $a$ and $b$ are the parameters. 
We use $\tau_l$ and $\tau_f$ to represent headways for the leading vehicle and the following vehicle respectively. In practice, $\tau_f$ is much smaller than $\tau_l$ due to small intra-platoon headways.

The controller at vertex $\nu_i$ instructs vehicle $k$ to travel with a constant speed $v_i^k$ over the coordinating zone $D_1$, and then its movements are decided by the IDM over the cruising zone $D_2$. The platoon formation in $D_2$ is determined by the headway of catching up $\tau_c$ and the intra-platoon headway $\tau_f$. When the headway of a vehicle is smaller than $\tau_c$, it will adjust its speed until its headway is less than $\tau_f$. In addition, the split process is defined by the headway $t_s$, which denotes the time until a vehicle that keeps a headway larger than $\tau_f$ from its leading vehicle is split off \cite{liang2015heavy}.

\subsection{Problem definition}
We consider a Markov decision process (MDP) for junction-level platooning coordination. Vehicle $k$ enters vertex $\nu_i$'s coordinating zone at $t^k_i$. We assume that $\{t_i^k: k=1,2,\ldots\}$ is a renewal process with a probability density function $g(x)$, and the inter-arrival time is $x_i^k = t_i^k - t_i^{k-1}$. The state is $s^k_i = [h^k_i \quad d^k \quad d^{k-1} \quad e^{k-1}_{ij}]$, where the predicted headway $h^k_i$ denotes the time used to catch up with the leading vehicle. $d^{k}$ is the destination of vehicle $k$, and $e^{k-1}_{ij}$ denotes the assigned edge for vehicle $k-1$ at $t^{k-1}_i$. The action is $a^k_i = [u^k_i \quad e^k_{ij}]$, where $e^k_{ij}$ denotes the assigned edge for vehicle $k$, and $u^k_i$ is the time reduction for vehicle $k$, which can be converted into the reference speed $v^k_i$ over the coordinating zone by $v^k_i = \frac{D_1}{D_1 / v_0 - u^k_i}$. The deterministic relationship between vehicle $k$ and vehicle $k-1$ is therefore, $h^k_i = x^k_i + u^{k-1}_i$, which indicates that current vehicle's predicted headway is determined by the time reduction of its leading vehicle. Furthermore, $h^k_i$ is larger than $x^k_i$ when the controller instructs vehicle $k-1$ to accelerate. Note that the speed over the coordinating zone $v^k_i$ is not affected by the IDM model.

Vehicle $k$ follows the routing decision from the controller at vertex $\nu_i$ and travels on edge $e^k_{ij}$ until it enters the coordinating zone of the adjacent vertex $\nu_j$ at time $t^k_j$, then detectors at vertex $\nu_j$'s coordinating zone collect the arrival time and fuel consumption of vehicle $k$. We consider the travel cost $c^k_{i}$ consisting of travel time $\Delta t^k_{ij} = t^k_j - t^k_i$
and fuel consumption $\Delta F^k_{ij}$ between vertex $\nu_i$ and vertex $\nu_j$. The total cost for vehicle $k$ on edge $e_{ij}$ is defined as
\begin{equation}
\begin{aligned}
\label{Eq: link_cost}
    c^k_{i} & = w_1 \Delta t^k_{ij} + w_2 \Delta F^k_{ij} \\
    & =w_1 \left(t^k_j - t^k_i \right) + w_2 \int_{t^k_i}^{t^k_j} (1- \eta \mathbbm{1}_{u_i^k = h_i^k}) r(v_t) dt,
\end{aligned}
\end{equation}
where $w_1$ is the value of time, and $w_2$ is the fuel price. The following vehicle in a platoon can save a fraction
$\eta \in (0,1)$ of the fuel consumption due to reduced air resistance, which only occurs when vehicle $k$ merges with its leading vehicle, i.e., $u_i^k = h_i^k$.
$r(v_t)$ is the fuel rate depending on the vehicle speed $v_t$.
Equation~(\ref{Eq: link_cost}) states the trade-off between benefits of platooning over $D_2$ and the increased cost due to acceleration over $D_1$.

The objective is to minimize travel cost in networks. At vertex $\nu_i$, $\mathbf{\mathcal{S}}_i$ denotes the state space, $\mathbf{\mathcal{A}}_i$ denotes the action space, and $\mathbf{\mathcal{C}}_i: \mathbf{\mathcal{S}}_i \times \mathbf{\mathcal{A}}_i \rightarrow \mathbb{R}$ is the travel cost on $e_{ij}$. When vehicle $k$ enters the coordinating zone of vertex $\nu_i$, the controller executes an action $a^k_i \in \mathbf{\mathcal{A}}_i$ according to the state $s^k_i \in \mathbf{\mathcal{S}}_i$, and receives the cost $c^k_{i} \in \mathbf{\mathcal{C}}_i$ when vehicle $k$ enters vertex $\nu_j$'s coordinating zone. We need to find the optimal policy $\pi_i^*: \mathbf{\mathcal{S}}_i \rightarrow \mathbf{\mathcal{A}}_i$ such that $a_i^k \sim \pi_i^*(\cdot | s_i^k)$ to minimize the accumulative cost from vertex $\nu_i$ to destinations.
Note that the optimality of $\pi_i^*$ relies on optimal policies of subsequent vertices. The network-wide policy is therefore $\pi^* = \left\{\pi_1^*, \pi_2^*, \ldots, \pi_N^* \right\}$.

%% file: sections/03_cascade.tex
\section{Coordinated platooning without routing}
\label{sec_cascade}

In this section, we discuss the platooning coordination without routing to analyze the interaction among controllers. It is difficult to find exact solutions due to the interdependencies among controllers' decisions. To address this, we propose two general approaches using approximate dynamic programming for suboptimal control in the context of a stochastic finite horizon problem. The numerical results show that the approach using approximation in the policy space can improve data-efficiency without loss of the coordination performance, which will be extended to the network-level coordination considering platooning and routing in Section~\ref{sec_network}.

\subsection{Coordination over a cascade of junctions}
\begin{figure}[!hbt]
  \centering
  \includegraphics[width=0.68\textwidth, trim=500 430 520 430,clip]{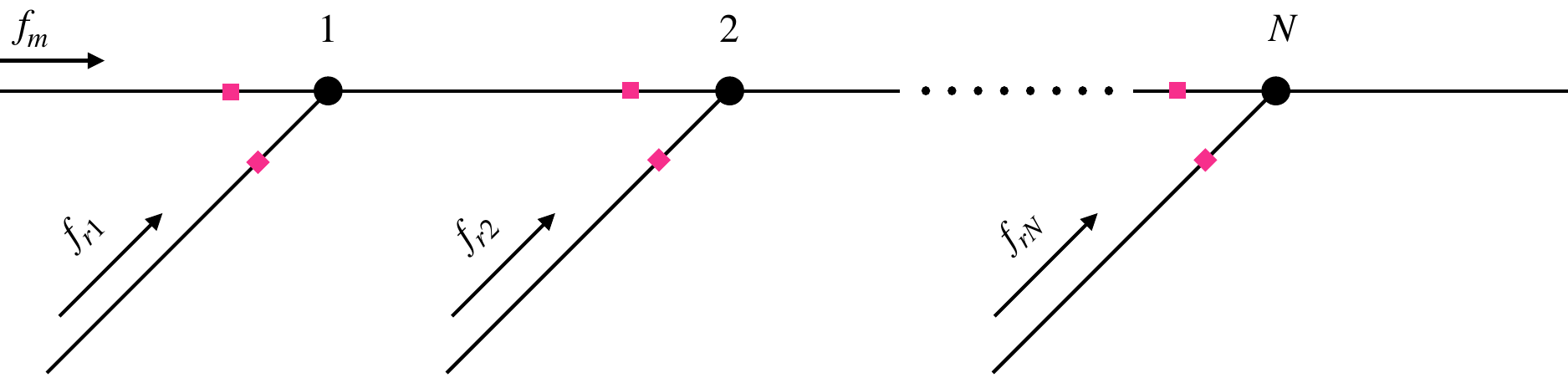}
  \caption{Coordinated platooning over a cascade of junctions.}\label{fig:cascade_shape}
\end{figure}

We study coordinated platooning over a cascade of junctions to analyze the interaction among controllers.
Fig.~\ref{fig:cascade_shape} presents the road segment with $N$ junctions. Each junction is equipped with a virtual controller, and red points denote the start of the coordinating zone of the mainline and on-ramps. When vehicle $k$ enters the coordinating zone of vertex $\nu_i$, $i=1,2,\ldots,N$, the controller at vertex $\nu_i$ determines whether vehicle $k$ merges with its leading vehicle $k-1$ based on $h_i^k$. The flow on the mainline is $f_m$, and the flow on the ramp of vertex $\nu_i$ is $f_{ri}$. Let $\mathbf{\mathcal{H}}_i$ denote the state space at vertex $\nu_i$, and $\mathbf{\mathcal{U}}_i$ is the action space. The optimal policy is $\mu_i^*: \mathbf{\mathcal{H}}_i \rightarrow \mathbf{\mathcal{U}}_i$ such that $u_i^k \sim \mu_i^*(\cdot | h_i^k)$.

We assume that the travel speed is related to the traffic density on the road segment, which in turn affects both travel time and fuel consumption. We propose two general approaches for coordinated platooning considering time-dependent traffic flows: (i) approximation in the value space, and (ii) approximation in the policy space. The first approach is data-driven and employs the framework of dynamic programming. Since we only consider the predicted headway $h_k$ as the decision variable, the polynomial regression is utilized to approximate the cost function. As for the second approach, we directly approximate the platooning policy with estimated travel times. In the following subsections, we will introduce both approaches and compare the numerical results obtained from each.


\subsection{Approximation in the value space}
Let $J_i^*(h_i^k, \xi_i^k)$ be the expected cost from vertex $\nu_i$ to the destination with the predicted headway $h_i^k$ and the action $\xi_i^k \in \mathcal{U}_i=\{0, 1\}$, where $\xi_i^k = 1$ denotes the merging case with $u_i^k = h_i^k$, while $\xi_i^k = 0$ represents that vehicle $k$ keeps its original speed $v_0$ over the coordinating zone. The expected cost $J_i^*(h_i^k, \xi_i^k)$ can be expressed as
\begin{align*}
J_i^*(h_i^k, \xi_i^k) = \mathbb{E} \left[c_{i}^k + c_{i+1}^k + \cdots + c_N^k \right],
\end{align*}
where $c_i^k, i = 1, 2, \ldots, N$, is the travel cost from vertex $\nu_i$ to vertex $\nu_{i+1}$. $J_i^*(h_i^k, \xi_i^k)$ can be estimated within the framework of dynamic programming,
\begin{align}
\label{Eq: Network Bellman Equation}
J_i^*(h_i^k, \xi_i^k) = \mathbb{E} \left[c_i^k + \min_{\xi_{i+1}^k}J_{i+1}^{*}(h_{i+1}^k, \xi_{i+1}^k)\right],
\end{align}
where $h_{i+1}^k$ is the predicted headway of vehicle $k$ 
arriving at vertex $\nu_{i+1}$, and $\xi_{i+1}^k$ is the action at vertex $\nu_{i+1}$. When vehicle $k$ enters the coordinating zone of vertex $\nu_{i+1}$, it returns the travel cost from vertex $\nu_i$ to vertex $\nu_{i+1}$, and the estimated cost from vertex $\nu_{i+1}$ to its destination.

When $\xi_i^k = 1$, we approximate $J_i^*(h_i^k, \xi_i^k)$ with a nonlinear polynomial function,
\begin{align}
\label{Eq: poly_reg}
J_i^*(h_i^k, \xi_i^k) \approx \beta_0 + \beta_1 h_i^k + 
\cdots + \beta_n (h_i^k)^{n} + \epsilon_i^k,
\end{align}
where $n$ is the degree, $\beta_0, \beta_1, \ldots, \beta_n$ are the parameters, and $\epsilon_i^k \sim N(0, \sigma^2)$ is the random error. Consider $l$ vehicles arriving at vertex $\nu_i$ before vehicle $k$ , i.e., vehicles $k-1, k-2, \ldots, k-l$, the observed state consisting of vehicle $k$ and previous $l$ vehicles can be expressed as the matrix notation,
\begin{align*}
    & \mathbf{H}_i^k=\left[\begin{matrix} 
    1  & h_i^k & (h_i^k)^2 & \cdots   & (h_i^k)^n \\
    1  & h_i^{k-1} & (h_i^{k-1})^2 & \cdots   & (h_i^{k-1})^n \\
    1  & h_i^{k-2} & (h_i^{k-2})^2 & \cdots   & (h_i^{k-2})^n \\
    \vdots   & \vdots  & \vdots    & \ddots   & \vdots \\
    1  & h_i^{k-l} & (h_i^{k-l})^2 & \cdots   & (h_i^{k-l})^n
  \end{matrix}\right].
\end{align*}
Note that vehicles $k-1, k-2, \ldots, k-l$ are all with the merging action. The corresponding cost can be denoted as,

\begin{align*}
    \mathbf{J}_{i}^{k} = \left[\begin{matrix} 
    J_i^*(h_i^k, \xi_i^k)  & J_i^*(h_i^{k-1}, \xi_i^{k-1}) & \cdots   & J_i^*(h_i^{k-l}, \xi_i^{k-l})
    \end{matrix}\right]^{\mathsf{T}}.
\end{align*}

Hence, the $l+1$ regression equations in the matrix form can be expressed by, 
\begin{align*}
    \mathbf{J}_{i}^k = \mathbf{H}_{i}^k \boldsymbol{\beta}_{i}^k + \boldsymbol{\epsilon}_{i}^k,
\end{align*}
where $\boldsymbol{\beta}_{i}^k = \left[\begin{matrix} \beta_0  & \beta_1 & \cdots   & \beta_n
\end{matrix}\right]^{\mathsf{T}}$ 
denotes the parameter term. $\boldsymbol{\epsilon}_{i}^k = \left[\begin{matrix} \epsilon_i^k  & \epsilon_i^{k-1} & \cdots   & \epsilon_i^{k-l}
\end{matrix}\right]^{\mathsf{T}}$ represents the error term. By utilizing the least squared error approach in the matrix form \cite{ostertagova2012modelling}, the estimated parameter term assuming $n < l+1$ is,
\begin{align*}
    \hat{\boldsymbol{\beta}}_{i}^k = \left((\mathbf{H}_{i}^k)^{\mathsf{T}} \mathbf{H}_{i}^k \right)^{-1} (\mathbf{H}_{i}^k)^{\mathsf{T}} \mathbf{J}_{i}^k.
\end{align*}

Then we can rewrite Equation~(\ref{Eq: Network Bellman Equation}) by,
\begin{align*}
J_i^*(h_i^k, \xi_i^k, \hat{\boldsymbol{\beta}}_i^k) = \mathbb{E} \left[c_i^k + \min_{\xi_{i+1}^k}J_{i+1}^{*}(h_{i+1}^k, \xi_{i+1}^k) \right],
\end{align*}
where $\min_{\xi_{i+1}^k}J_{i+1}^{*}(h_{i+1}^k, \xi_{i+1}^k)$ also consists of the merging action and non-merging action for vehicle $k$.

When $\xi_i^k = 0$, $J_i^*(h_i^k, \xi_i^k)$ is not related to $h_i^k$ since the travel cost does not depend on the predicted headway. We use the average cost of previous $y+1$ vehicles to represent the estimated cost,
\begin{align*}
    J_i^*(h_i^k, \xi_i^k) \approx \frac{1}{y+1} \sum_{j=0}^{y} \left[c_i^{k-j} + \min_{\xi_{i+1}^{k-j}}J_{i+1}^{*}(h_{i+1}^{k-j}, \xi_{i+1}^{k-j}) \right],
\end{align*}
where the previous $y+1$ vehicles are all with non-merging action. The proposed algorithm for parametric cost approximation with polynomials is shown in Algorithm \ref{algorithm:rl_polynomial}.

\begin{algorithm}[htbp]
  \caption{Parametric cost approximation with polynomials.}
  \label{algorithm:rl_polynomial}
  \begin{algorithmic}[1]
    \Require
      Initialization of $J_i^*(h_i, 0)$;
      Initialization of $\boldsymbol{\beta}_{i}$;
      The polynomial degree $n$;
      Number of merging vehicles $l > n - 1$;
      Number of non-merging vehicles $y$.
      
    \Ensure
      Optimal platooning policy $\mu_i^*$;
    \State Vehicle $k$ enters the coordinating zone of vertex $\nu_{i+1}$
    \State $\xi_{i+1}^k \gets \argmin_{\xi_{i+1}^k} J_{i+1}^*(h_{i+1}^k, \xi_{i+1}^k)$
    \State $\hat{J}_{i+1}^k \gets \min_{\xi_{i+1}^{k}}J_{i+1}^{*}(h_{i+1}^{k}, \xi_{i+1}^{k})$
    \State Update vertex $\nu_i$ with $c_i^k$ and the estimated cost
    \If{$\xi_i^k = 0$}
        \State $\tilde{J}_{i+1}^k \gets \min_{\xi_{i+1}^{k-j}}J_{i+1}^{*}(h_{i+1}^{k-j}, \xi_{i+1}^{k-j})$
        \State $J_i^*(h_i^k, \xi_i^k) \gets \frac{1}{y+1} \sum_{j=0}^{y} \left[c_i^{k-j} + \tilde{J}_{i+1}^k \right]$
    \Else
        \State $J_i^*(h_i^k, \xi_i^k, \boldsymbol{\beta}_{i}^k) \gets c_i^k + \hat{J}_{i+1}^k$
        \State $\hat{\boldsymbol{\beta}}_{i}^k \gets \left((\mathbf{H}_{i}^k)^{\mathsf{T}} \mathbf{H}_{i}^k \right)^{-1} (\mathbf{H}_{i}^k)^{\mathsf{T}} \mathbf{J}_{i}^k$
    \EndIf
    \State Output the optimal policy $\mu_i^*$ such that
    \If{$\argmin_{\xi_{i}} J_{i}^*(h_{i}, \xi_{i}) = 1$}
        \State $\mu_i^*(h_i) = h_i$
    \Else
        \State $\mu_i^*(h_i) = 0$
    \EndIf
\end{algorithmic}
\end{algorithm}

\subsection{Approximation in the policy space}
Typically, we can use the relative cost instead of absolute values in Equation~(\ref{Eq: link_cost}),
\begin{align*}
    \label{Equation: reward_function}
    RC_i^k = -w_1 u_i^k + w_2 (\Delta F_1 - \Delta F_2),
\end{align*}
where $\Delta F_1$ is the increased fuel during $D_1$ due to speed adaption,
\begin{align}
    \Delta F_1 = \alpha D_1 \left( \frac{D_1}{D_1/v_0 - u_i^k} \right)^2 - \alpha D_1 {v_0}^2,
\end{align}
where $\alpha$ is the coefficient of increased fuel with speed and travel distance. $\Delta
 F_2 = \eta \phi D_2$ arises when $u_i^k = h_i^k$, where $\phi$ denotes the fuel efficiency. When vehicle $k$ is not instructed to merge with its leading vehicle, i.e., $u_i^k < h_i^k$, we have $\Delta F_2 = 0$. The relative cost for vehicle $k$ is,
\begin{align}
\label{Equation: total_cost}
    RC_i^k = \begin{cases}
    -w_1 u_i^k + w_2 \left(\Delta F_1  - \eta \phi D_2 \right) & u_i^k = h_i^k, \\
    -w_1 u_i^k + w_2  \Delta F_1 & u_i^k < h_i^k.
     \end{cases}
\end{align}

The optimal policy $\mu^*(h)$ for the junction-level coordination under general arrival processes is a threshold-based policy such that
\begin{align}
\label{Equation: threshold_policy}
    \mu^*(h)=\begin{cases}
    h & h \leq \theta,\\
    c & h > \theta,
    \end{cases}
\end{align}
where $\theta$ is the threshold, and $c$ is the time reduction \cite{xiong2021optimizing} . The $k$th vehicle would merge with its leading vehicle $k-1$ if the predicted headway $h_k$ is less than the threshold $\theta$; otherwise vehicle $k$ should slightly decelerate in anticipation with constant time reduction $c$ in anticipation of further subsequent vehicles to merge. Furthermore,
$\theta$ and $c$ in Equation~(\ref{Equation: threshold_policy}) can be computed by solving the following system of equations under Poisson arrivals with $g(x) = \lambda e^{- \lambda x}$,
\begin{align}
\label{Equation: Poisson_solutions}
    \begin{cases}
        Z - e^{\lambda (1-\gamma) \theta} \biggl(\int_{c}^{\theta} e^{- \lambda (1-\gamma) t}  \left(G^{'}(t) - \lambda G(t) \right) \mathrm{d}t \\
         + \left(Z + G(0) \right) e^{- \lambda (1-\gamma) c} \biggl) = 0 , \\
        G(\theta) + \gamma Z - Z = 0, \\
         G^{'}(c) - \lambda G(c) + \lambda (1 - \gamma) \left( Z + G(0) \right) = 0,
    \end{cases}
\end{align}
where $\lambda$ is the arrival rate, $\gamma$ is the discount factor, and $G^{'}(u)$ is the derivative of $G(u)$,
\begin{equation*}
\begin{aligned}
G(u)  = & w_1 u + w_2 \biggl( \alpha D_1 {v_0}^2 - \alpha D_1 \left( \frac{D_1}{D_1/v_0 - u} \right)^2 \\
     & + \eta \phi D_2 \biggl),
\end{aligned}
\end{equation*}
where $G(u)$ denotes the merging cost when $u < D_1/{v_0}$. When we consider multiple junctions in Fig.~\ref{fig:cascade_shape}, the Poisson process can be utilized to approximate vehicle arrivals at vertex $\nu_i$ since headways within a platoon are much smaller than inter-arrival times.

\begin{figure}[!hbt]
  \centering
  \includegraphics[width=0.60\textwidth, trim=180 70 180 70,clip]{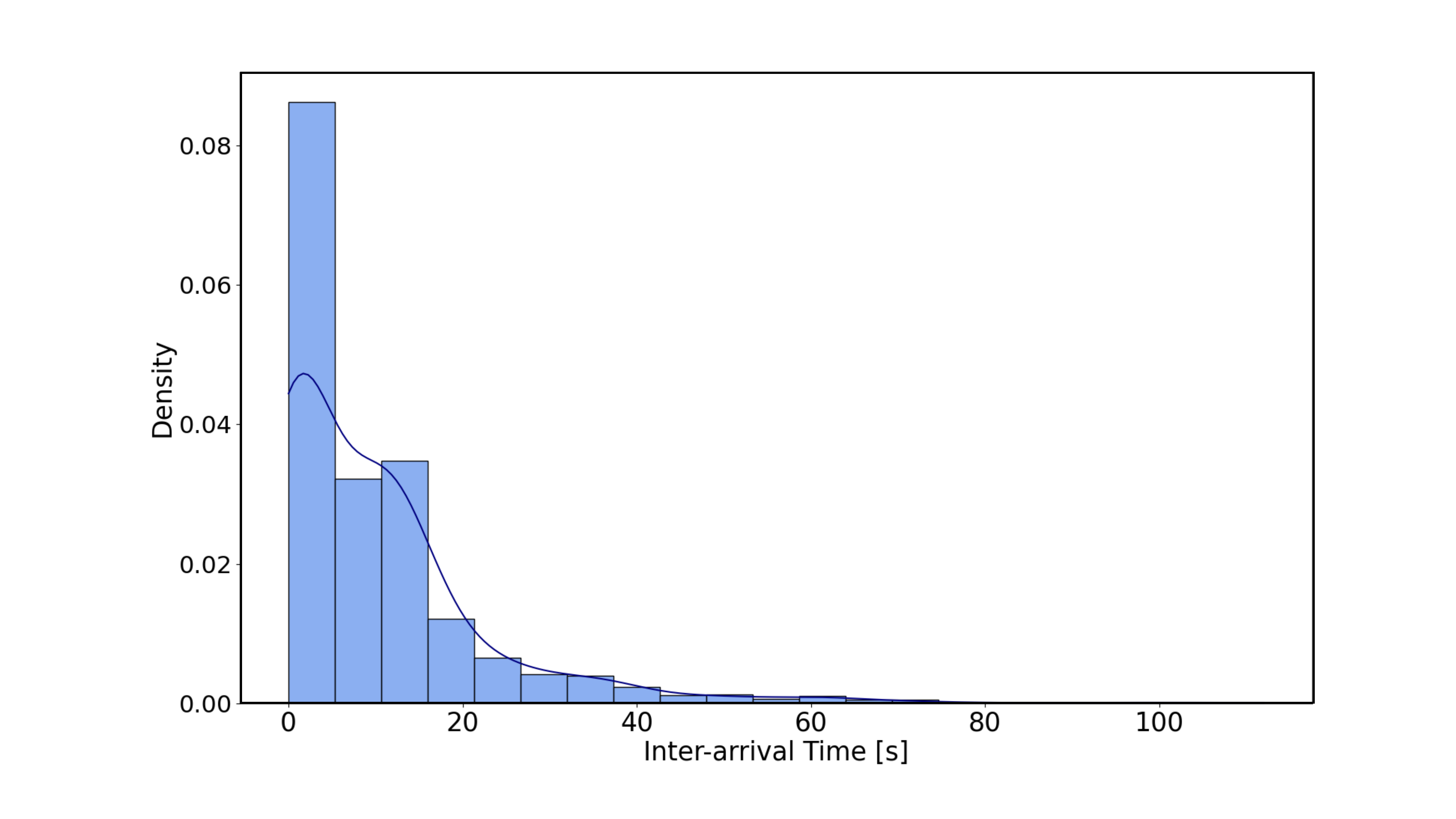}
  \caption{Empirical density of CAV's inter-arrival time at vertex $\nu_2$.}\label{fig:poisson_process}
\end{figure}

We have evaluated vehicle arrivals for vertex $\nu_2$ in Fig.~\ref{fig:cascade_shape}. $f_m, f_{r1}, f_{r2}$ are all assumed to be Poisson arrivals with arrival rates $\lambda_m = \lambda_{r1} = 108$ veh/h, $\lambda_{r2} = 180$ veh/h, and the number of vehicle arrivals $N_{m} = N_{r1} = 500$ veh, $N_{r2} = 1000$ veh. The headways within a platoon $h_0$ are assumed to be $0.5$ s, and the nominal speed $v_0 = 24$ m/s. The edge between vertex $\nu_1$ and vertex $\nu_2$ is $31$ km consisting of $D_2 = 30$ km and $D_1 = 1$ km. The threshold-based policy in Equation~(\ref{Equation: threshold_policy}) is utilized at vertex $\nu_1$ for platooning coordination with $D_2 = 30$ km.
Fig.~\ref{fig:poisson_process} shows the empirical density of inter-arrival time at vertex $\nu_2$. The majority of inter-arrival time is the headway within a platoon, and the probability density function shows that the Poisson process can be utilized to approximate vehicle arrivals at vertex $\nu_2$. Since $h_0$ is much smaller than $x_i^k$ and accounts for the majority of inter-arrival times, Poisson process can be used to approximate vehicle arrivals at each vertex when $f_m, f_{r1}, f_{r2}, \ldots f_{rN}$ all follow Poisson arrivals, thus Equation~(\ref{Equation: Poisson_solutions}) can be used to calculate the threshold-based policy at each vertex.

Furthermore, $\eta \phi D_2$ in Equation~(\ref{Equation: reward_function}) denotes the fuel reduction over the cruising zone. However, it assumes that $D_2$ is constant at a single-junction, which is not realistic in time-dependent networks. We utilize the framework of dynamic programming to update the travel time from vertex $\nu_i$ to its destination,
\begin{align}
   \Delta T_i^{k} = \chi \left(U_i^k + T_{i+1}^k -T_i^k\right),
\end{align}
where $T_i^k$ is vehicle $k$'s estimated travel time from vertex $\nu_i$ to the destination, $U_i^k$ is the real travel time from vertex $\nu_i$ to $\nu_{i+1}$, and $T_{i+1}^k$ is the estimated travel time from vertex $\nu_{i+1}$ to the destination. When vehicle $k$ enters the coordinating zone of vertex $\nu_{i+1}$, it returns $U_i^k$ and $T_{i+1}^k$ to update $T_i^k$, and $\chi$ is the update rate. 
We employ this approach to update travel times in time-dependent networks since $U_i^k$ is the real travel time.
Additionally, we utilize this approach to estimate the cruising distance by $\hat{D}_i^{k} = \frac{r(\overline{v}_i^k) T_i^{k}}{\phi}$,
where $\overline{v}_i^k = D_2 / T_i^{k}$ is average speed over $D_2$. 
To estimate the arrival rate $\lambda_i^k$ when vehicle $k$ enters the coordinating zone of $\nu_i$, we utilize the $M$-step discounted previous headways, which helps account for flow fluctuation,
\begin{align}
\label{Equation: arrival_rate}
    \hat\lambda_i^k=\Big[(1-\psi)\sum_{m=0}^{M-1}\psi^m x_i^{k-m}\Big]^{-1},
\end{align}
where $\psi$ is the discount factor of inter-arrival times. Therefore, we can use Equation~(\ref{Equation: Poisson_solutions}) to determine the coordination policy with estimated parameters,
\begin{align}
\label{Equation: threshold_policy_parameter}
    \mu_i^*(h_i^k, \hat{D}_i^{k}, \hat\lambda_i^k)=\begin{cases}
    h_i^k & h_i^k \leq \theta_i^k,\\
    c_i^k & h_i^k > \theta_i^k,
    \end{cases}
\end{align}
where $\theta_i^k$ and $c_i^k$ can be computed using $\hat{D}_i^{k}$ and $\hat\lambda_i^k$.
The proposed algorithm for approximation in the threshold-based policy is shown in Algorithm~\ref{algorithm:threshold_based policy}.

\begin{algorithm}[htbp]
  \caption{Approximation in the threshold-based policy.}
  \label{algorithm:threshold_based policy}
  \begin{algorithmic}[1]
    \Require
        The cruising distance $D_2$ at vertex $\nu_{i}$;
        The fuel rate function $r(v)$;
        The fuel efficiency $\phi$; 
        The reward function $G(u)$;
        The discount factor of inter-arrival times $\psi$;
        Number of discounted steps $M$;
    \Ensure
      Optimal platooning policy $\mu_i^*$;

    \State Vehicle $k$ enters the coordinating zone of vertex $\nu_{i}$
    \State $\hat\lambda_{i}^k \gets \Big[(1-\psi)\sum_{m=0}^{M-1}\psi^m x_{i}^{k-m}\Big]^{-1}$
    \State $\hat{D}_i^{k} \gets \frac{r(\overline{v}_{i}^k) T_{i}^{k}}{\phi}$
    \State Determine $\theta_{i}^k$ and $c_{i}^k$ with $\hat\lambda_{i}^k$ and $\hat{D}_i^{k}$ using Equation~(\ref{Equation: Poisson_solutions})
    \If{$h_{i}^k \geq \theta_{i}^k$}
        \State $\mu_i^*(h_i^k) = c_{i}^k$
    \Else
        \State $\mu_i^*(h_i^k) = h_{i}^k$
    \EndIf
        
    \State Update the previous vertex $\nu_{i-1}$ with $U_{i-1}^k$ and $T_{i}^k$

    $\Delta T_{i-1}^{k} \gets \chi \left(U_{i-1}^k + T_{i}^k -T_{i-1}^k\right)$
    
\end{algorithmic}
\end{algorithm}


\subsection{Numerical results}
\begin{figure}[hbt]
  \centering
  \includegraphics[width=0.65\textwidth, trim=80 340 70 300,clip]{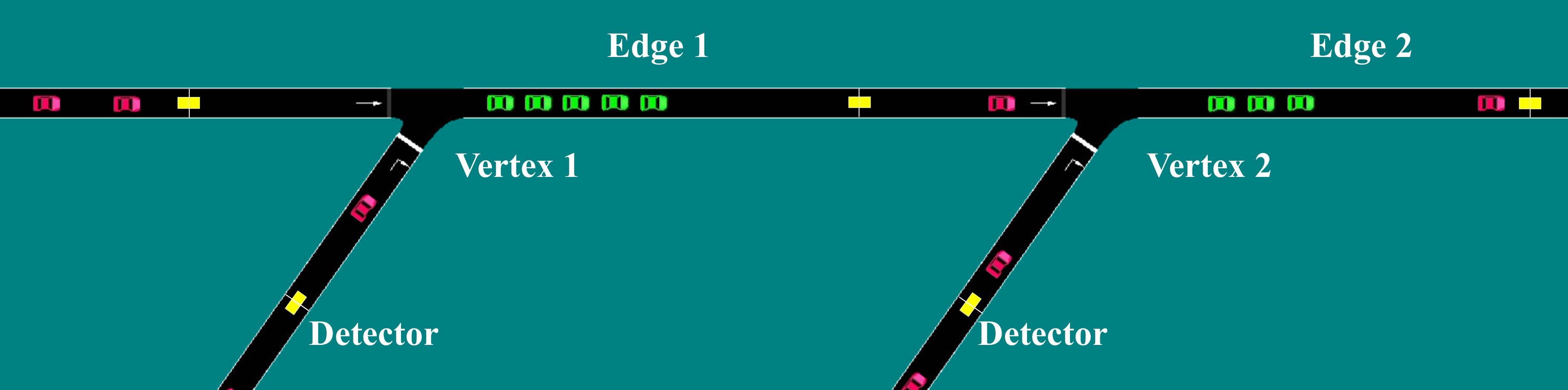}
  \caption{SUMO platform for platooning over two consecutive junctions.}\label{fig:sumo_cascade}
\end{figure}

We consider the platooning coordination over two consecutive junctions in the Simulation of Urban MObility (SUMO) \cite{SUMO2018}, as shown in Fig.~\ref{fig:sumo_cascade}. The coordinating distance $D_1$ is set to be $1$ km for both junctions, and the cruising distances for vertices $1$ and $2$ are both set to be $30$ km. When a CAV enters the coordinating zone, the controller at the vertex determines whether the current CAV merges with the CAV ahead based on the predicted headway. Green vehicles in Fig.~\ref{fig:sumo_cascade} denote CAVs in a platoon, red vehicles are CAVs traveling alone, and yellow points represent detectors collecting vehicle arrival times. The desired headway for the leading vehicle in a platoon $\tau_l$ is $7.5$ s, and the following CAVs' movements are determined by the platoon management function \textit{Simpla} in SUMO to keep a small headway.

When a CAV enters the coordinating zone, it returns the travel cost on the edge and the expected travel cost from the current vertex to its destination. The cost in the simulation consists of fuel consumption and travel time. The fuel rate $f$ (L/s) is related to the travel speed $v$ (m/s), and can be derived using the empirical data in SUMO: $f = 3.51 \times 10^{-7} v^3 + 4.07 \times 10^{-4} v$. The fraction of fuel saving compared with traveling alone $\eta$ can be up to $21\%$ for smaller headways within a platoon. We follow the setting in \cite{larson2015distributed} and assumes a $10\%$ fuel reduction rate. The update rate $\chi$ is set to be $1.0$ since we neglect the normalization effect during the travel time estimation. Other nominal values are shown in Table~\ref{table:simulation_nominal_parameters}. We consider traffic flows consisting of CAVs and non-CAVs. The Greenshield's model is utilized to simulate the macroscopic speed-density relationship, $v_e= v_0 \left(1.0 - k_e / k_j \right)$, where $k_e$ is the density on the edge, and $k_j$ is the jam density. Let $k_c = {k_j}/{2}$ be the critical density. The speed $v_e$ is used to limit the CAV or platoon speed on the edge. Non-CAVs are used to simulate the congestion effect, and the penetration rate of CAVs $\zeta$ is set to be $1/6$.

\begin{table}[htbp]
  \caption{Nominal parameters in the simulation.}\label{table:simulation_nominal_parameters}
  \begin{center}
    \begin{tabular}{ccc}
    \hline
    Parameter                         & Notation   & Value                \\
    \hline
    Fuel efficiency                   & $\phi$     & $32.2 \ L/100 km$    \\
    Nominal speed                     & $v_0$      & $24 \ m/s$           \\
    Value of time                     & $w_1$      & $25.8 \ \$/hour $    \\
    Fuel price                        & $w_2$      & $0.868 \ \$ / L$     \\
    Coordinating distance             & $D_1$      & $1 \ km$             \\
    Fraction of fuel saving           & $\eta$     & $0.1$                \\
    Discount factor in the policy     & $\gamma$   & $0.9$                \\
    Discount factor of arrival times  & $\psi$     & $0.9$                \\
    Number of steps                   & $M$        & $50$                 \\
    Update rate                       & $\chi$     & $1.0$                \\
    \hline
    \end{tabular}
  \end{center}
\end{table}

\begin{figure}[!hbt]
  \centering
  \includegraphics[width=0.6\textwidth, trim=320 120 380 210,clip]{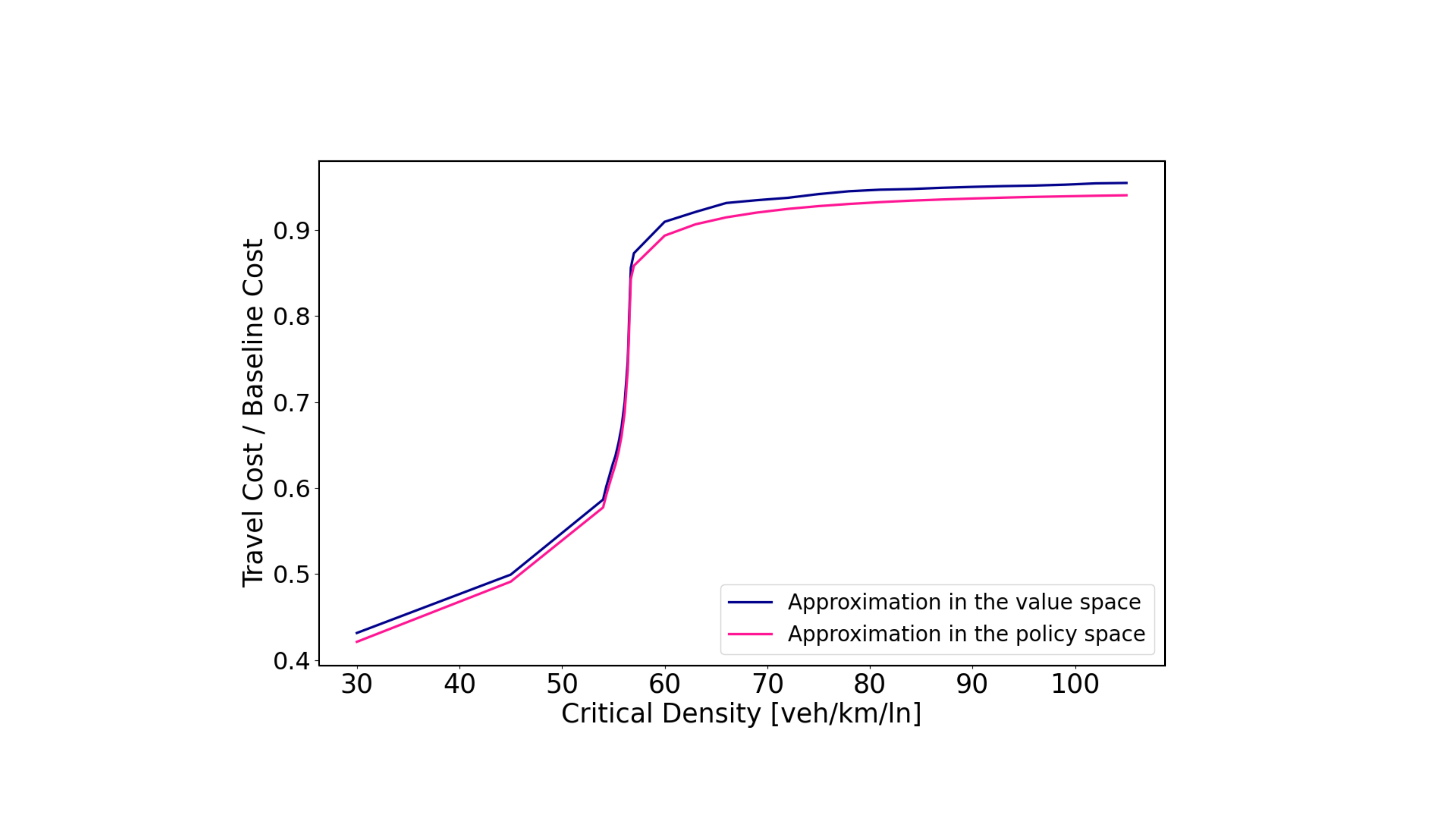}
  \caption{Ratio of the travel cost to the baseline cost with the critical density.}\label{fig:cost_ratio_density}
\end{figure}

We generate 2000 CAVs to evaluate the coordination performance in the SUMO platform. We compare the performance between parametric cost approximation with polynomials and approximation in the threshold-based policy. Additionally, we consider the case without platooning as the baseline. The results in Fig.~\ref{fig:cost_ratio_density} show the ratio of the travel cost to the baseline cost with the critical density. The smaller the critical density, the more likely the road will be congested. We use the critical density as the variable instead of the traffic flow since larger flow will increase the number of simulated vehicles, thus increasing the computation burden. The results demonstrate that using approximation in the threshold-based policy yields better performance than parametric cost approximation with polynomials. Both curves indicate that the ratio increases as the critical density increases, which shows that platooning will have more benefits when the highway is congested. CAVs traveling in platoons will lead to more density reduction, thus reducing travel time and fuel consumption.

In Fig.~\ref{fig:cost_ratio_density}, we observe that when the critical density reaches $30$ veh/km/ln (indicating heavily congested conditions), platooning yields a significant travel cost reduction of over $50\%$ compared to the scenario without platooning. Moreover, as the critical density exceeds $70$ veh/km/ln, we can still achieve a notable reduction in travel cost of nearly $10\%$. It is worth noting that the sensitivity of the cost ratio diminishes in density intervals above $60$ veh/km/ln or below $50$ veh/km/ln. This reduced sensitivity arises because the travel cost and the baseline cost exhibit minimal changes during periods of heavy congestion or free flow traffic.

%% file: sections/04_network.tex
\section{Coordinated platooning and adaptive routing}
\label{sec_network}

In this section, we delve into the study of coordination within networks, encompassing coordinated platooning and adaptive routing at each junction. 
To address the complexity of the problem, we decouple the action space by prioritizing routing decisions based on travel time estimation. Subsequently, we employ the aforementioned policy approximation to determine speed profiles, considering parameters such as travel times and cruising distances.
The results in SUMO demonstrate the superior performance of our approach compared to conventional methods.
Additionally, we asses the resilience of our approach in dynamically changing networks, affirming its ability to maintain efficient platooning operations.

\subsection{The approximate dynamic programming approach}
Platooning coordination in networks consists of platooning and routing at each vertex. We need to determine the action $a_i^k = [u_i^k \quad e_{ij}^k]$ based on the state $s_i^k = [h_i^k \quad d^k \quad d^{k-1} \quad e_{ij}^{k-1}]$ when vehicle $k$ enters the coordinating zone of vertex $\nu_i$. To simplify the structure of the problem, we decouple the action space and prioritize the routing decision $e_{ij}^k$ since most drivers will not drive any more time than necessary \cite{larson2015distributed}. 
Our approach involves estimating travel times within the network using a dynamic programming framework, allowing for the determination of the optimal path. Subsequently, the platooning coordination is determined using the approximation in the threshold-based policy, as described in Section~\ref{sec_cascade}.
It is important to note that the estimated travel times can be utilized to calculate parameters within the threshold-based policy, facilitating efficient platooning coordination.


Let $T_i(d^k,j)$ denote the time that vertex $\nu_i$ estimates it takes to travel from $\nu_i$ to the destination $d^k$ for vehicle $k$ by way of vertex $\nu_i$'s adjacent vertex $\nu_j$. Similar to the approach for packet routing in the communication networks \cite{boyan1994packet}, the dynamic programming framework for updating travel times is shown as
\begin{align}
\label{Equation: rl_network}
    \Delta T_i(d^k,j) = \chi \left( U_{ij}^k + \min_{z \in \mathcal{N}_j} T_j(d^k,z) - T_i(d^k,j) \right),
\end{align}
where $U_{ij}^k$ is the travel time from vertex $\nu_i$ to vertex $\nu_j$ for vehicle $k$. $\mathcal{N}_j$ denotes adjacent vertices of vertex $\nu_j$. When vehicle $k$ enters the coordinating zone of vertex $\nu_j$, it returns $U_{ij}^k$ and vertex $\nu_j$'s estimate for the time remaining in the trip $\min_{z \in \mathcal{N}_j} T_j(d^k,z)$, which can be used to update $T_i(d^k,j)$. The routing decision $e_{ij}^k$ is determined based on the adjacent vertex with minimum $T_i(d^k,j)$. Note that $T_i(d^k,j)$ is updated in a large table instead of a continuous function.

\begin{figure}[hbt]
  \centering
  \includegraphics[width=0.56\textwidth, trim=380 230 420 250,clip]{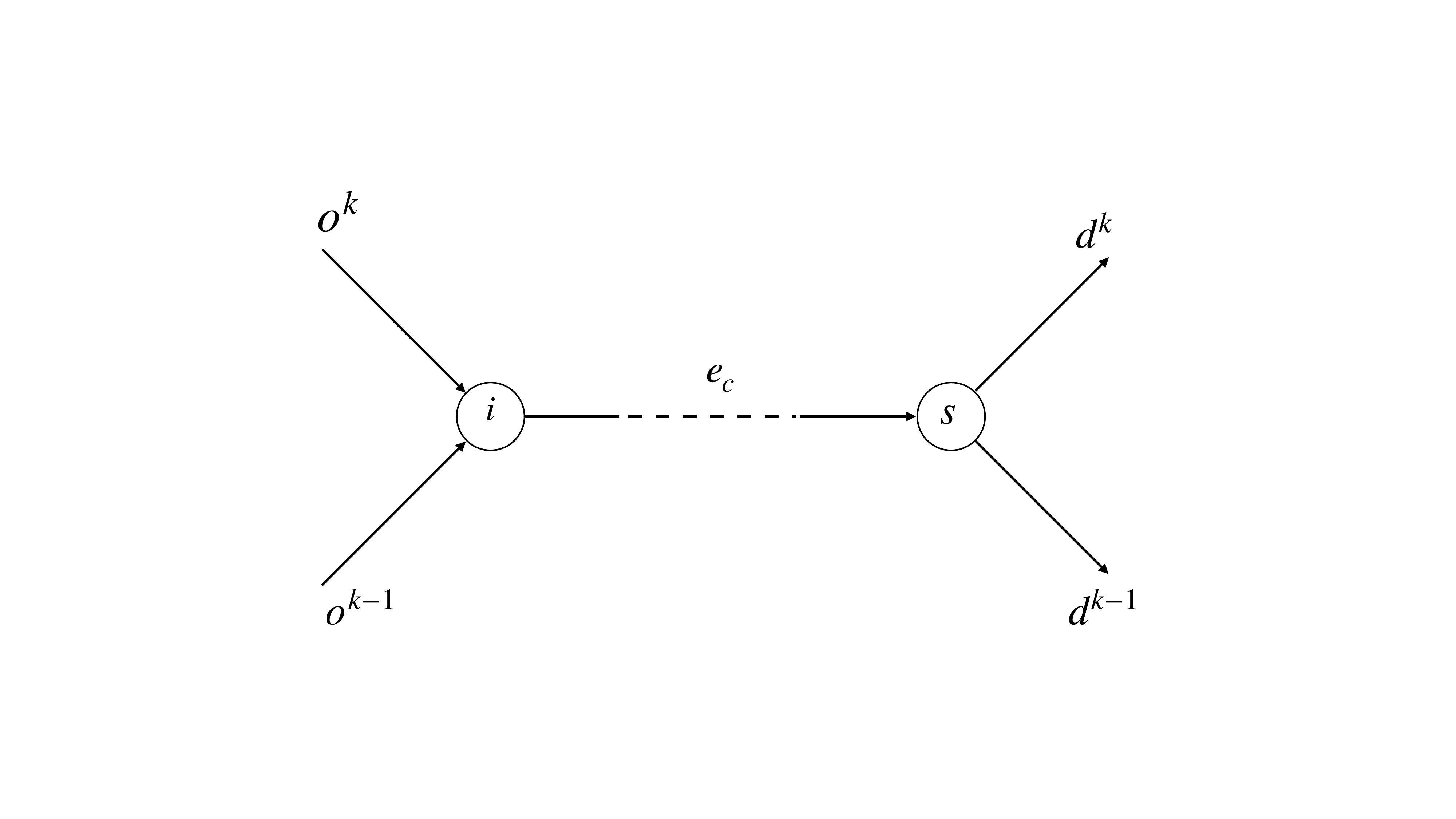}
  \caption{The platooning exists when vehicle $k$ and vehicle $k-1$ share a common path.}\label{fig:network_platooning}
\end{figure}

When vehicle $k$ and vehicle $k-1$ share a common path, the controller at the junction determines the coordination process using the approximation in the threshold-based policy. Fig.~\ref{fig:network_platooning} shows the platooning formation and split for vehicles with different destinations. Vehicle $k$ travels from the origin $o^k$ to the destination $d^k$, and vehicle $k-1$ travels from the origin $o^{k-1}$ to the destination $d^{k-1}$. The merging decision is made at vertex $\nu_i$, and the split junction is vertex $\nu_s$. Vehicle $k$ and vehicle $k-1$ share a common path $e_c$, which leads to the fuel reduction. The estimated cruising distance $\hat{D}_{e_c}$ can be expressed as
\begin{align}
    \hat{D}_{e_c} = \frac{1}{\phi} \left( r(\overline{v}_i) T_i^{k} - r(\overline{v}_s) T_s^{k} \right),
\end{align}
where $T_i^{k}$ is the estimated travel time from vertex $\nu_i$ to its destination $d^k$, and $T_s^{k}$ is the estimated travel time from vertex $\nu_s$ to $d^k$. $\overline{v}_i$ denotes the average speed from vertex $\nu_i$ to $d^k$, and $\overline{v}_s$ is the average speed from vertex $\nu_s$ to $d^k$.

$\hat{D}_{e_c}$ can be used to replace $D_2$ in $\Delta F_2 = \eta \phi D_2$ since the platooning only occurs on $e_c$. Furthermore, the platooning policy at vertex $\nu_i$ can be approximated with the threshold-based policy,

\begin{align}
\label{Equation: threshold_policy_network_parameter}
    u_i^k=\begin{cases}
    h_i^k & h_i^k \leq \theta_i^k,\\
    c_i^k & h_i^k > \theta_i^k,
    \end{cases}
\end{align}
where $\theta_i^k$ and $c_i^k$ can be calculated using Equation~(\ref{Equation: Poisson_solutions}) using $\hat{D}_{e_c}$ and $\hat{\lambda}_i^k$.
The proposed algorithm for coordinated platooning and adaptive routing is shown in Algorithm~\ref{algorithm:network_platooning}. When vehicle $k$ enters the coordinating zone of vertex $\nu_i$, the controller at vertex $\nu_i$ calculates the estimated arrival rate $\hat\lambda_{i}^k$. We can use $T_i(d^k, j)$ to determine the selected edge $e_{ij}^k$ and the split vertex $\nu_s$.
When vehicle $k$ and vehicle $k-1$ do not share a common path, the time reduction $u_i^k = c_i^k$, i.e., vehicle $k$ should slightly decelerate with $c_i^k$ to form a platoon with its following vehicle. Note that $c_i^k$ is calculated using the estimated cruising distance $\hat{D}_i^{k}$ from vertex $\nu_i$ to the destination. When vehicle $k$ and vehicle $k-1$ share a common path, the system of equations in Equation~(\ref{Equation: Poisson_solutions}) can be utilized to calculate $\theta_{i}^k$ and $c_{i}^k$ in the threshold-based policy.
After determining the routing and platooning at vertex $\nu_i$, $T_p(d^k,i)$ at the previous vertex $\nu_p$ is updated using the dynamic programming framework in Equation~(\ref{Equation: rl_network}).

\begin{algorithm}[htbp]
  \caption{The approximate dynamic programming for coordinated platooning and adaptive routing.}
  \label{algorithm:network_platooning}
  \begin{algorithmic}[1]
    \Require
        Vehicle $k$'s destination $d^k$;
        Vehicle $(k-1)$'s destination $d^{k-1}$;
        The fuel rate function $r(v)$;
        The fuel efficiency $\phi$;
        The reward function $G(u)$;
        The discount factor of arrival times $\psi$;
        Number of discounted steps $M$;
    \Ensure
      Optimal edge $e_{ij}^k$ and the time reduction $u_i^k$;

    \State Vehicle $k$ enters the coordinating zone of vertex $\nu_{i}$
    \State $\hat\lambda_{i}^k \gets \Big[(1-\psi)\sum_{m=0}^{M-1}\psi^m x_{i}^{k-m}\Big]^{-1}$
    \State Determine  $e_{ij}^k$ and $\nu_s$ using $T_i(d^k, j)$
    
    \If{$\nu_s = \nu_i$}
        \State $\hat{D}_i^{k} \gets \frac{r(\overline{v}_{i}^k) \hat{T}_{i}^{k}}{\phi}$
        \State Determine $\theta_{i}^k$ and $c_{i}^k$ with $\hat\lambda_{i}^k$ and $\hat{D}_i^{k}$
        \State $u_{i}^k = c_{i}^k$
    \Else
        \State $\hat{D}_{e_c} \gets \frac{1}{\phi} \left( r(\overline{v}_i) T_i^{k} - r(\overline{v}_s) T_s^{k} \right)$
        \State Determine $\theta_{i}^k$ and $c_{i}^k$ with $\hat\lambda_{i}^k$ and $\hat{D}_{e_c}$

        \If{$h_{i}^k \geq \theta_{i}^k$}
            \State $u_{i}^k = c_{i}^k$
        \Else
            \State $u_{i}^k = h_{i}^k$
        \EndIf
    \EndIf
        
    \State Update $T_p(d^k,i)$ at the previous vertex $\nu_p$
    \State $\Delta T_p(d^k,i) \gets \chi \left( U_{pi}^k + \min_{z \in \mathcal{N}_i} T_i(d^k,z) - T_p(d^k,i) \right)$
    
\end{algorithmic}
\end{algorithm}


\subsection{Numerical Results}
\begin{figure}[htp]
\centering

{\includegraphics[width=0.65\textwidth, trim=0 0 0 0,clip]{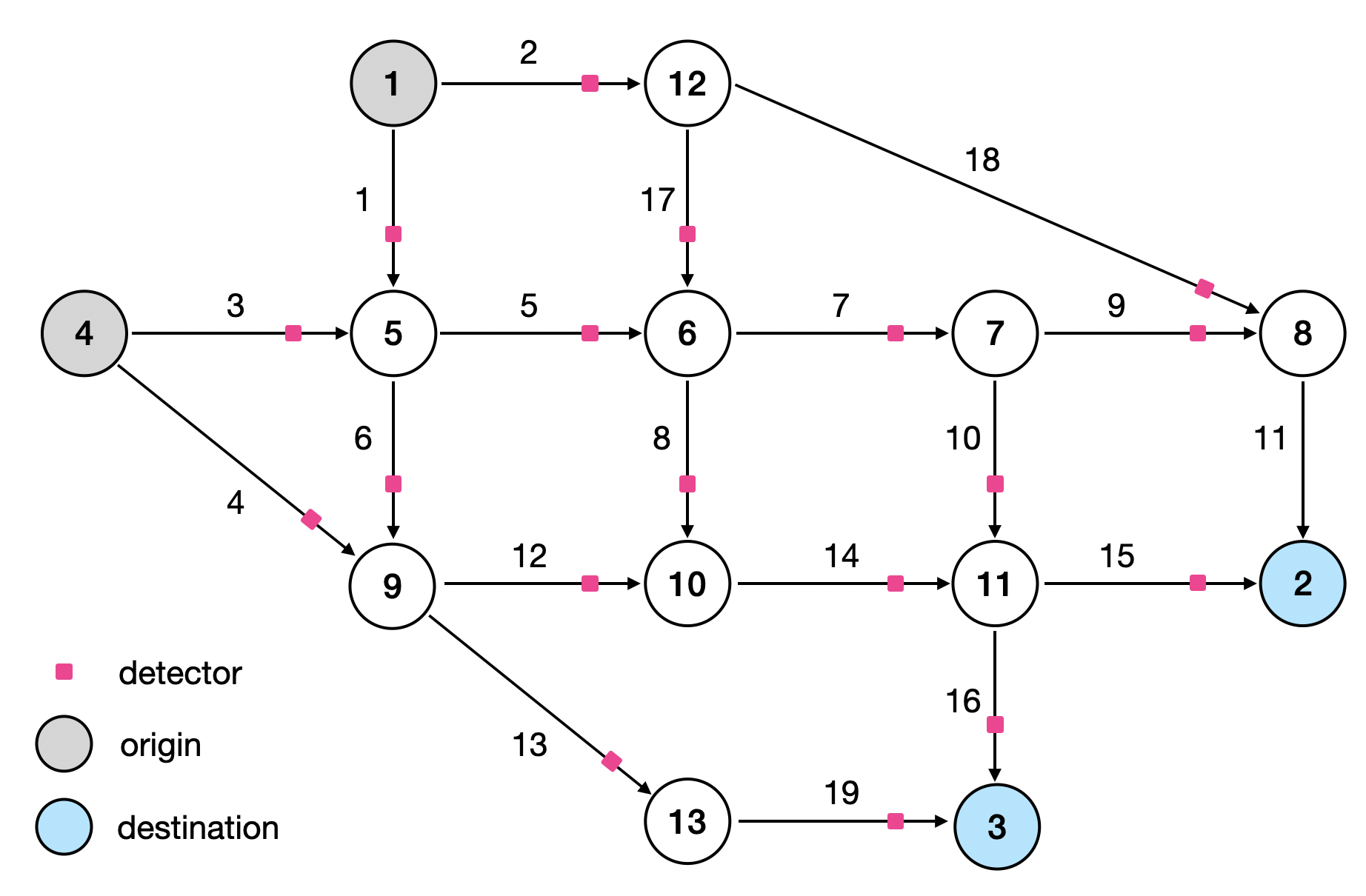}}
\caption{Platooning coordination in the Nguyen-Dupuis network.}
\label{fig:sumo_network}
\end{figure}

We evaluate our approach in the Nguyen-Dupuis network shown in Fig.~\ref{fig:sumo_network}. To simulate the microscopic vehicle dynamics, we employ the IDM car-following model, while the macroscopic Greenshield's model captures the congestion effect resulting from the mixed-autonomy traffic.
The network comprises 13 vertices and 19 edges, consisting of 2 origins and 2 destinations. 
The red points in Fig.~\ref{fig:sumo_network} represent the detectors strategically placed to collect trip information, which can be shared through the network with the communication infrastructure. When a vehicle enters the coordinating zone of a junction, the controller will determine the appropriate routes and speed profiles. It will then proceed to update the time table at the previous junction, ensuring coordinated platooning operations throughout the network.

We generate 500 CAVs for each origin-destination (O-D) pair with the Poisson arrival process, where the arrival rate is set to be $216$ veh/hr. The penetration rate of CAVs $\zeta$ is set to be $1/10$. To improve the computation efficiency in SUMO, $D_1$ is set to be $500$ m and the distance of all edges is $2000$ m. Other parameters are the same as those in Table~\ref{table:simulation_nominal_parameters}. We compare our approach with other methods under different critical densities:

1. Baseline: each CAV travels alone its shortest path and no platoon is formed at the junction. CAVs will keep the original speed during the coordinating zone.

2. Policy A: we refer to the coordinated platooning framework in \cite{larson2015distributed}. Each vehicle is assumed to travel alone its shortest path. A platoon is formed only when the merging benefits outweigh the costs of traveling alone. The platooning decision is made using the policy in \cite{xiong2019analysis}, where only acceleration is allowed in the coordinating zone.

3. Policy B: our proposed approximate dynamic programming framework for coordinated platooning and adaptive routing.

\begin{figure}[hbt]
  \centering
  \includegraphics[width=0.55\textwidth, trim=330 120 380 210,clip]{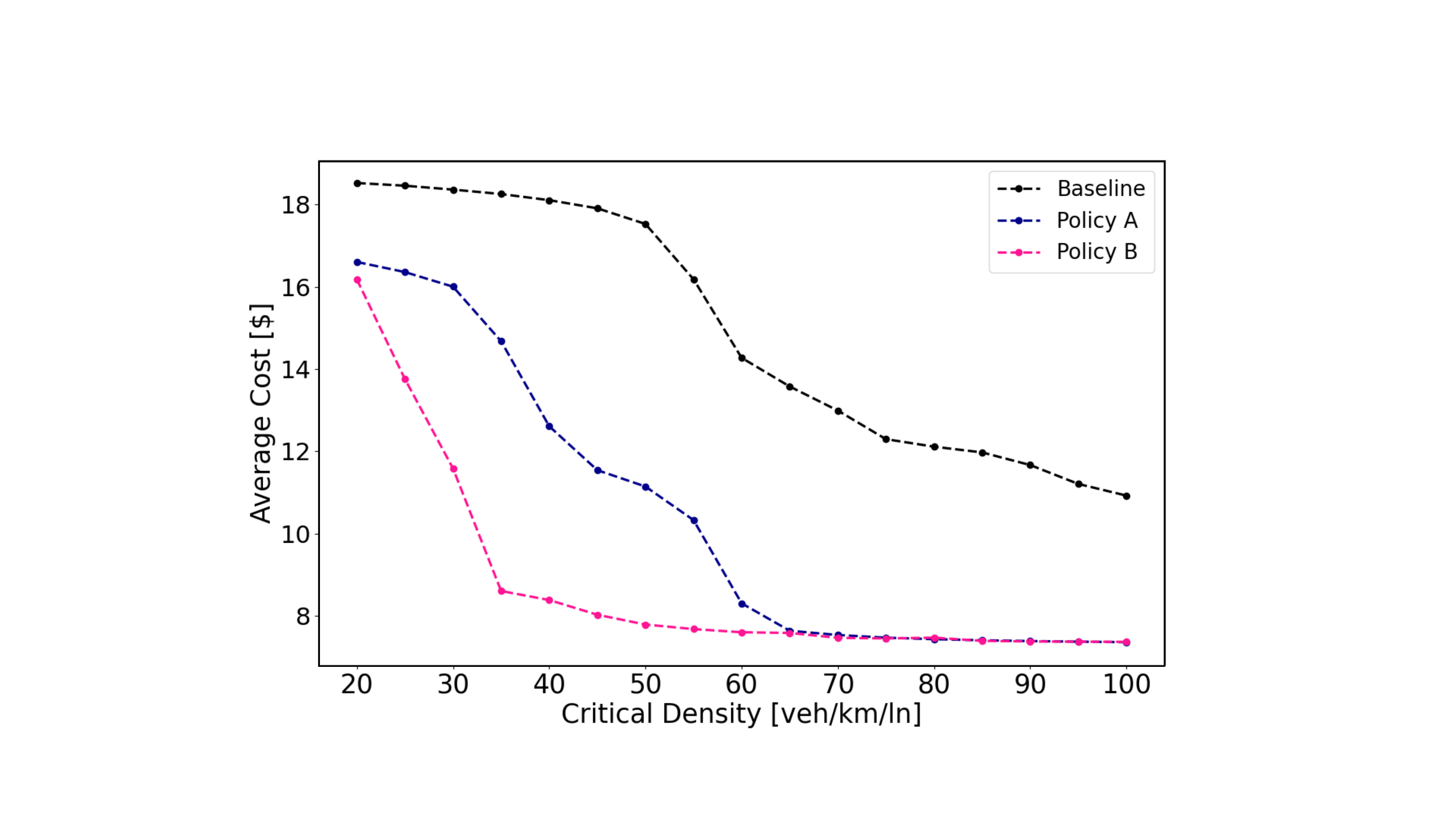}
  \caption{Performance comparison among different policies.}\label{fig:network_performance}
\end{figure}

Fig.~\ref{fig:network_performance} shows the relationship between the average cost and the critical density, where the average cost is obtained by dividing the total cost by the number of CAVs. The results show that the average cost will increase when the road is prone to congestion with smaller critical densities. Both Policy A and policy B can reduce travel cost compared with the baseline. Furthermore, our proposed policy B yields better performance than policy A, especially when the critical density is less than $60$ veh/km/ln. If the highway is congested, the dynamic programming framework can be used for adaptive routing with travel time estimation instead of traveling alone the shortest distance. The threshold-based policy in Equation~(\ref{Equation: threshold_policy_network_parameter}) yields better performance than the policy in \cite{xiong2019analysis} since deceleration is considered in the merging process. Compared with policy A, our proposed method can reduce the travel cost by up to $40\%$ when the critical density approximates $35$ veh/km/ln. Policy A and policy B achieve almost the same performance when the critical density approximates $20$ veh/km/ln since the road is heavily congested even though our method can adaptively choose the optimal path. When the critical density is greater than $70$ veh/lm/ln, policy A and policy B can also have similar performance since the highway is under free flow traffic.

\begin{figure}[hbt]
  \centering
  \includegraphics[width=0.55\textwidth, trim=330 120 380 210,clip]{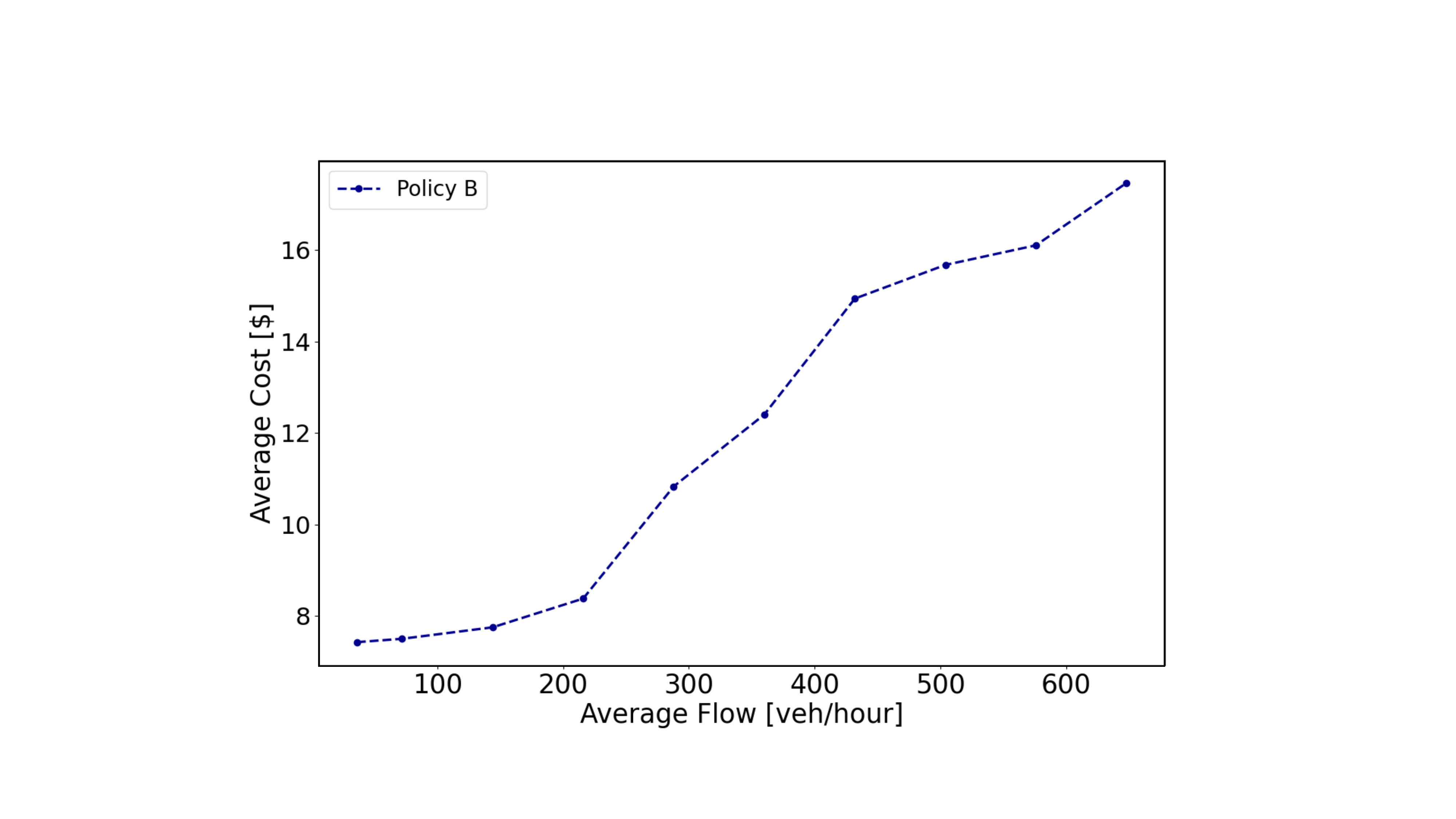}
  \caption{Average cost with different origin-destination demands.}\label{fig:cost_with_flow}
\end{figure}

We also evaluate our approach under different O-D demands, which are represented by traffic flows between origins and destinations. The results in Fig.~\ref{fig:cost_with_flow} indicate that the average cost increases with the average flow since more vehicles in the network will decrease the travel speed, thus increasing the travel time and fuel consumption. Note that the curve slope diminishes when the flow is less than $200$ veh/hr, resulting from the reduced density with platoons.

\subsection{Resiliency in dynamically changing networks}
\begin{figure*}[htb]
  \centering
  \subfigure[$T=700$ s.]
  {\label{fig:colorbar_700}
  \includegraphics[width=0.48\columnwidth,trim=30 60 30 90,clip]{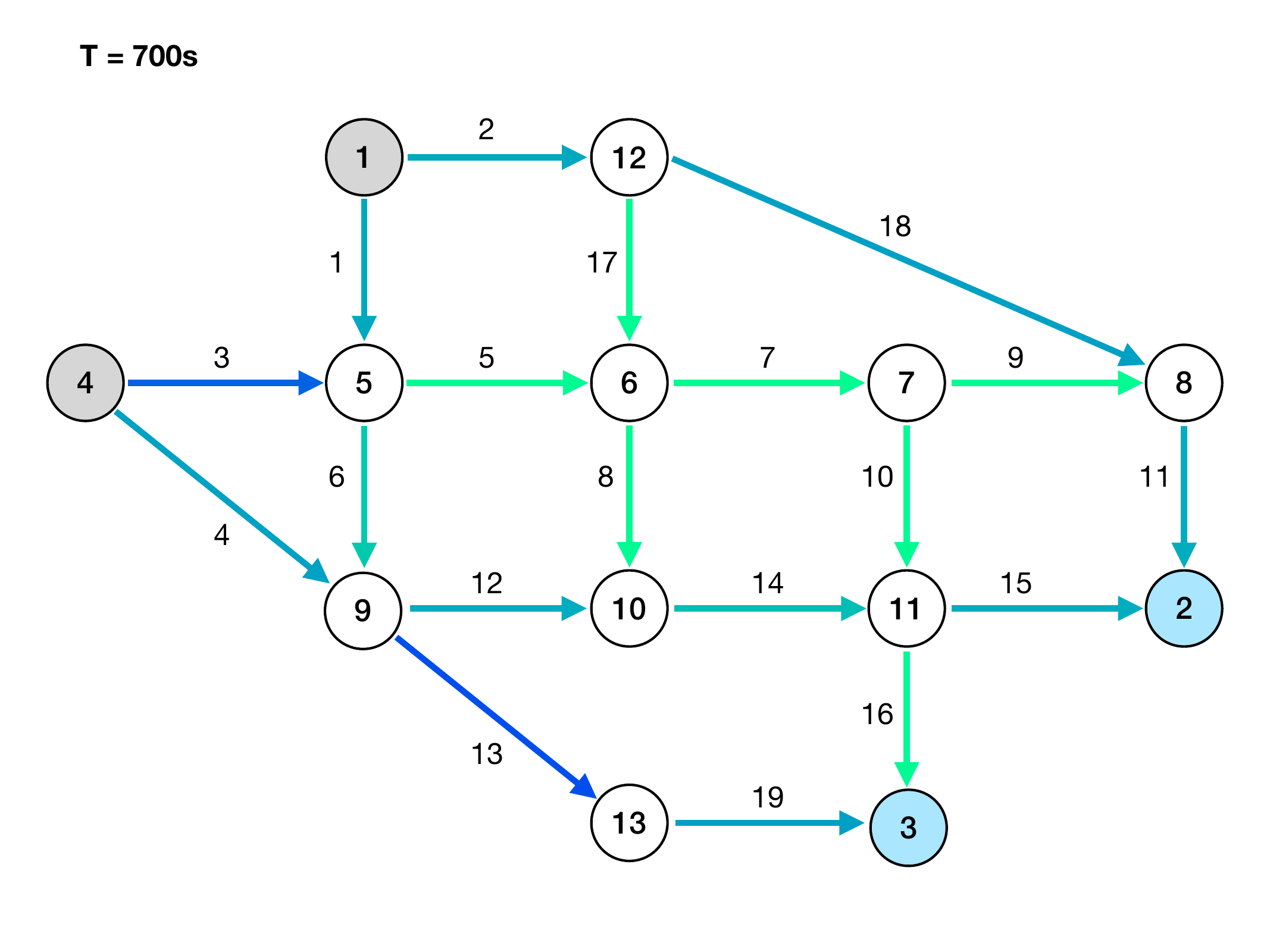} \label{fig:network_colorbar_700}
  }
  \subfigure[$T=1600$ s.]
  {\label{fig:colorbar_1600}
  \includegraphics[width=0.48\columnwidth,trim=30 60 30 90,clip]{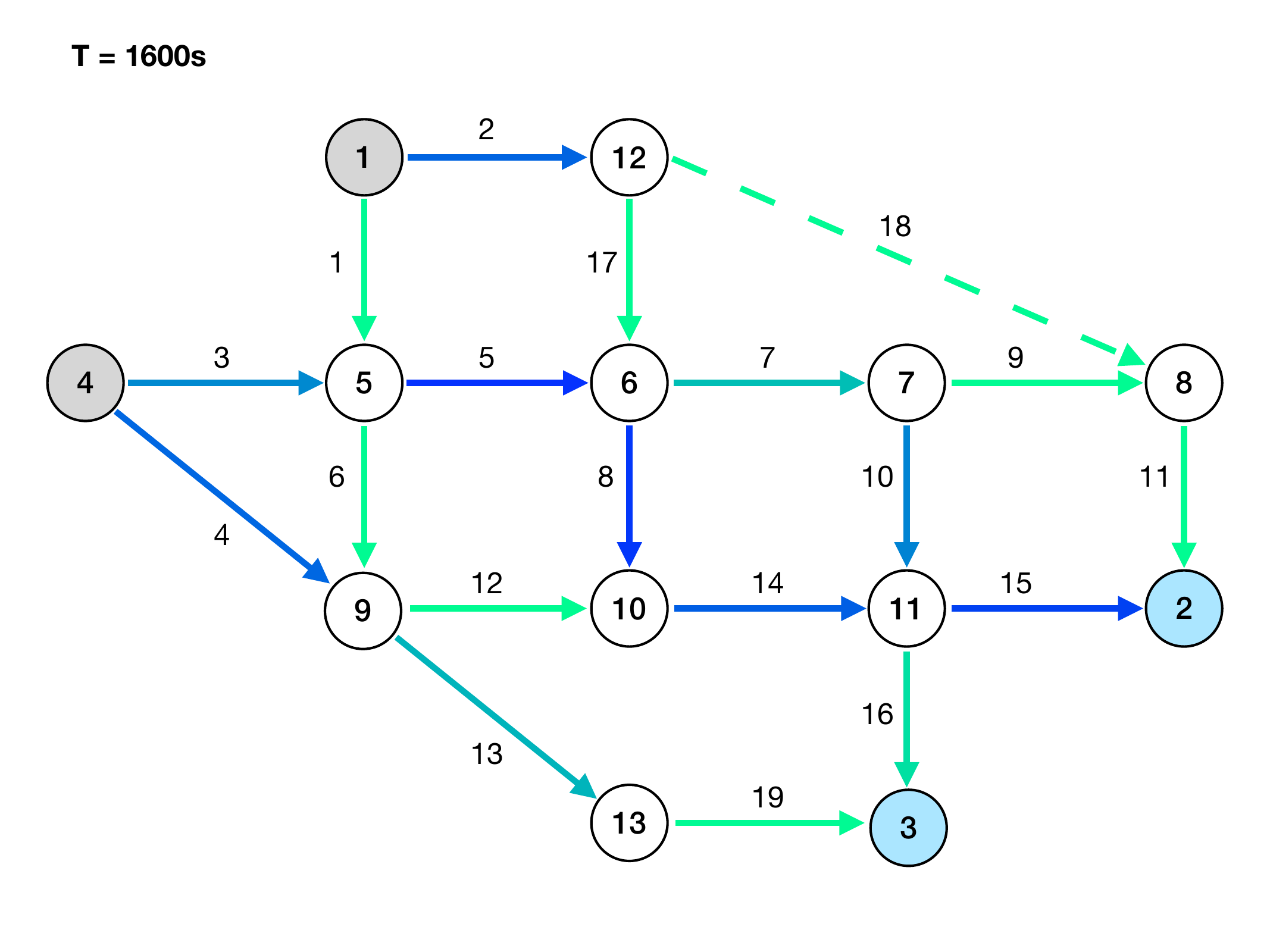}
  \label{fig:network_colorbar_1600}
  }
  \caption{The average speed for all edges before and after the disconnection of edge $18$.} \label{fig:network_colorbar}
\end{figure*}

We evaluate the resilience of our approach in dynamically changing networks by disconnecting edges during simulation. The platooning simulation is implemented in the SUMO platform with our proposed approach. The total simulation time is $4000$ s. We manually disconnect edge $18$ during $1000-2000$ s and resume the traffic after that. Fig.~\ref{fig:network_colorbar_700} displays the average speed for all edges at $700$ s before the disconnection. Light green color indicates a high average speed and low edge density, as observed in edges $5$ and $7$. In contrast, dark blue color represents greater density, as seen in edges $1$ and $18$.
The optimal path between origin $1$ and destination $2$ includes the edges $2-18-11$. 
Fig.~\ref{fig:network_colorbar_1600} depicts the average speed for all edges at $1600$ s. After the disconnection, more vehicles choose to travel on edges $5$, $8$ and $14$, indicating that our dynamic programming approach for travel time estimation can react to such changes and is able to continue routing traffic efficiently.

\begin{figure*}[hbt]
  \centering
  \includegraphics[width=0.96\textwidth, trim=350 150 370 100,clip]{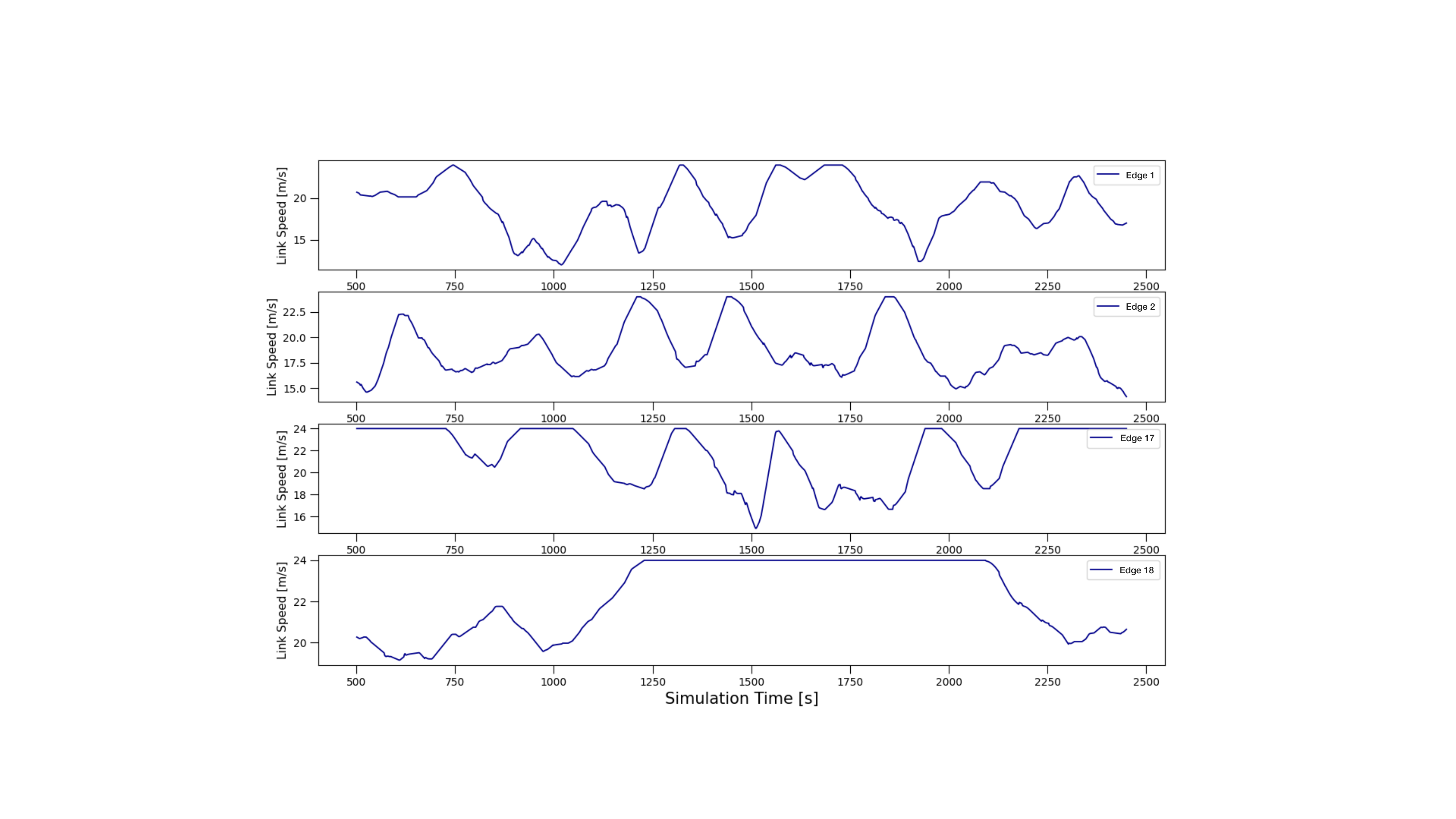}
  \caption{The average speed for edges $1$, $2$, $17$, and $18$.}\label{fig:link_average_speed}
\end{figure*}

In Fig.~\ref{fig:link_average_speed}, we specifically focus on edges $1$, $2$, $17$ and $18$ to illustrate the average speed variations throughout the simulation.
Edge $18$ is disconnected during $1000-2000$ s, resulting in an average speed of $24$ m/s during this period.
Note that edges $1$ and $2$ serve as alternate routes for vehicles to reach their destinations, and their average speed exhibit opposite directions. 
During the disconnection period, both edge $2$ and $17$ show similar patterns but with a delay of $120$ s, as edge $17$ serves as  the only downstream option for vehicles on edge $2$ after the disconnection. This observation demonstrates the adaptability and effectiveness of our approach in reacting to disruptions within the network.

\begin{figure*}[!htb]
  \centering
  \subfigure[Threshold value at vertex $8$.]
  {\label{fig:threshold_with_flow_node8}
  \includegraphics[width=0.44\columnwidth,trim=320 130 300 210,clip]{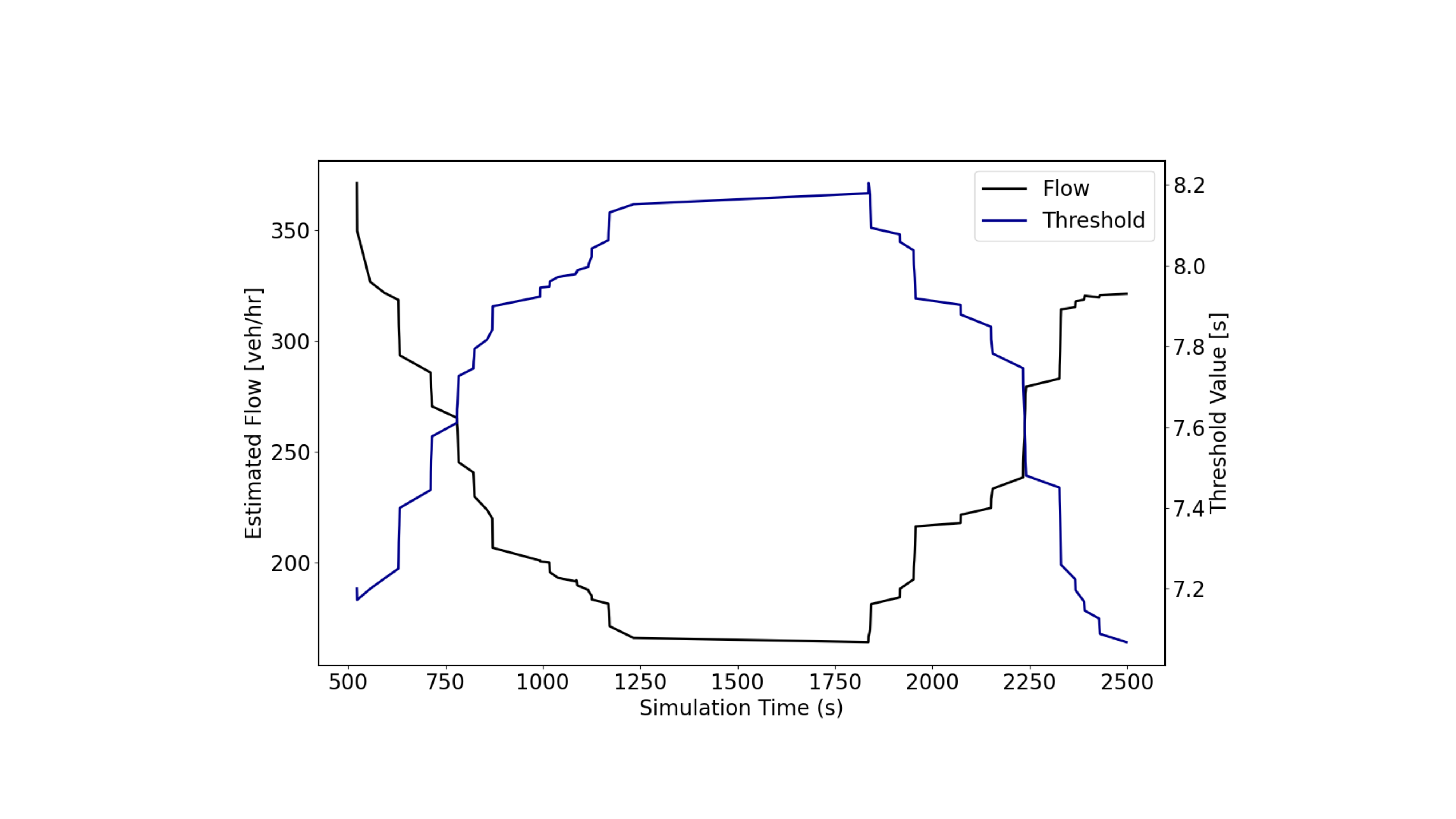}
  }
  \subfigure[C value at vertex $8$.]
  {\label{fig:C_with_flow_node8}
  \includegraphics[width=0.45\columnwidth,trim=320 130 260 210,clip]{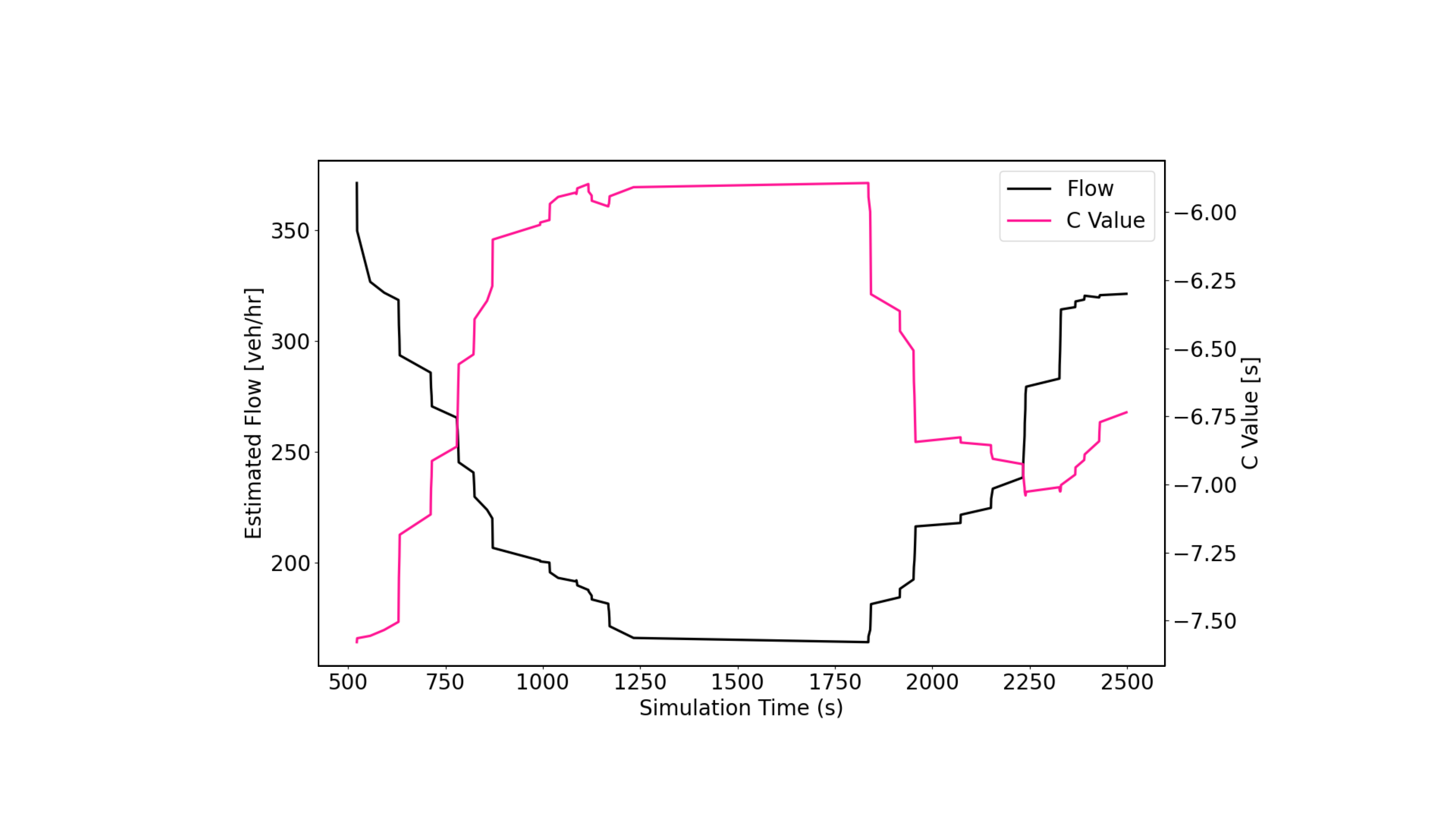}
  }
  \subfigure[Threshold value at vertex $5$.]
  {\label{fig:threshold_with_flow_node5}
  \includegraphics[width=0.45\columnwidth,trim=320 130 280 210,clip]{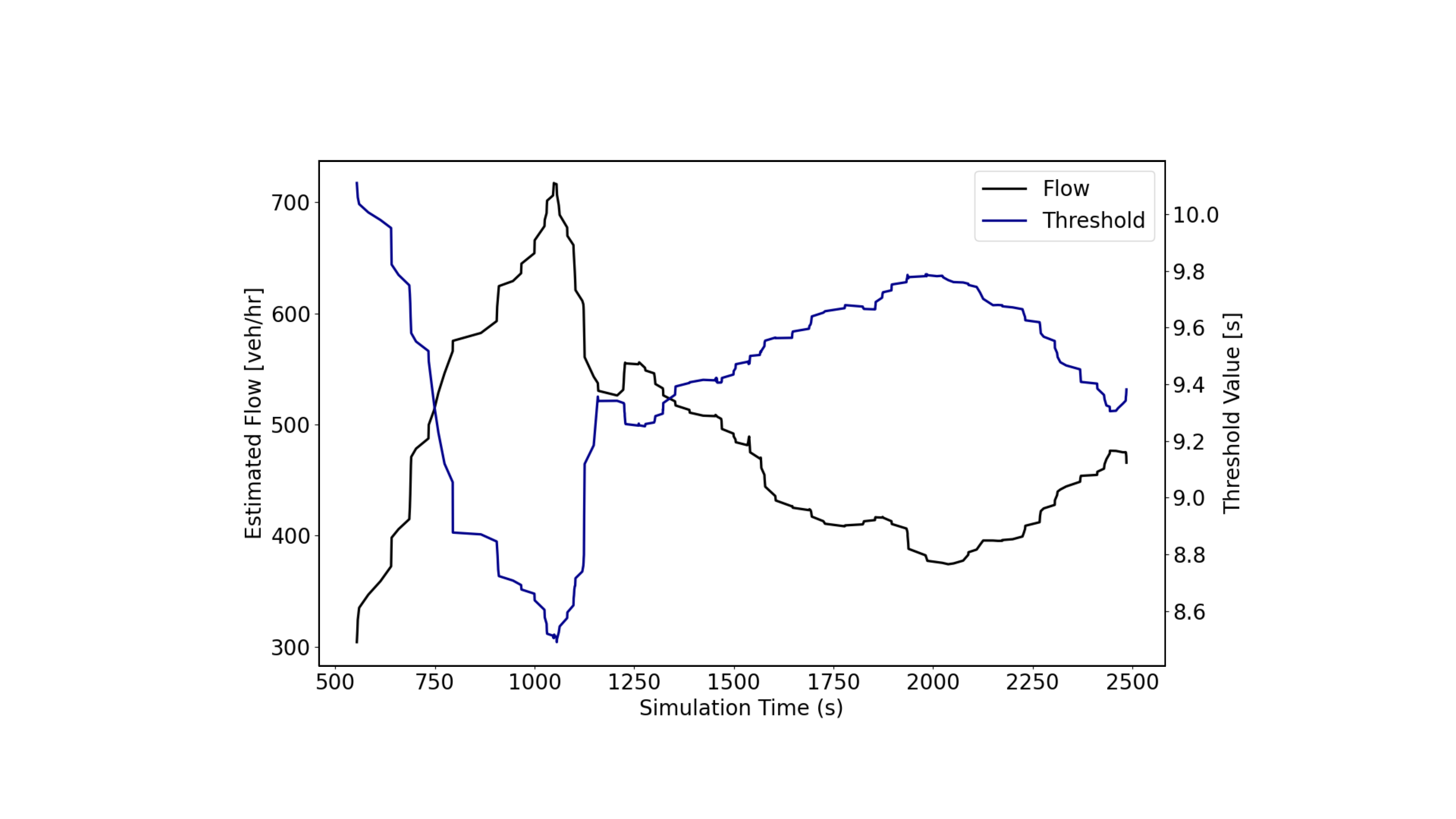}
  }
  \subfigure[C value at vertex $5$.]
  {\label{fig:C_with_flow_node5}
  \includegraphics[width=0.45\columnwidth,trim=320 50 280 210,clip]{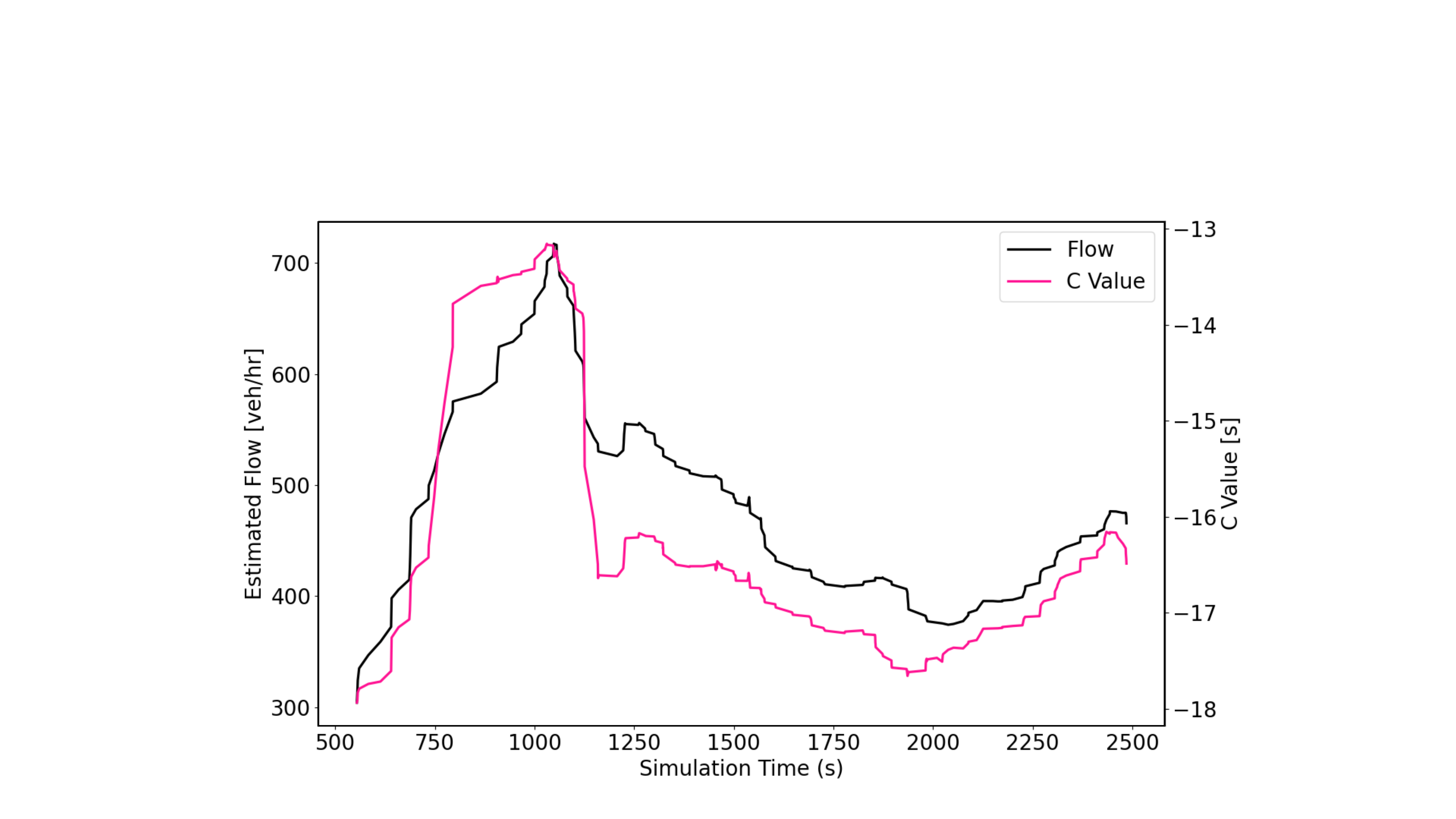}
  }
  
  \caption{Threshold and C values with estimated flows for vertices $8$ and $5$.} \label{fig:threshold_C_with_flow}
\end{figure*}

To demonstrate the real-time platooning strategy under varying traffic flows, we select two junctions within the network.
Fig.~\ref{fig:threshold_C_with_flow} shows threshold and C values with estimated traffic flows at vertices $8$ and $5$. 
The plots in Fig.~\ref{fig:threshold_with_flow_node8} and Fig.~\ref{fig:threshold_with_flow_node5} indicate that the threshold curve and the flow curve exhibit opposite directions of change.
Furthermore, Fig.~\ref{fig:C_with_flow_node8} reveals that the C value curve and the flow curve at vertex $8$ also show opposite trends.
Conversely, Fig.~\ref{fig:C_with_flow_node5} demonstrates that the C value curve and the flow curve at vertex $5$ exhibit similar trends, resulting from the longer cruising distance at that vertex.
Additionally, C values at vertex $5$ are much smaller than those at vertex $8$, resulting in a larger gap between the threshold and the C value at vertex $5$.
Consequently, vertex $5$ allows for more platooning due to benefits associated with longer cruising distances.

%% file: sections/05_conclusion.tex
\section{Concluding Remarks}
\label{sec_conclude}

In this paper, we have investigated platooning in time-dependent networks using an approximate dynamic programming framework.
By formulating a Markov decision process at each junction, our study focuses on minimizing travel time and fuel consumption.
We have analyzed platooning coordination without routing and proposed two innovative approaches based on approximate dynamic programming to address interdependencies among controllers' decisions. The results over a cascade of junctions have demonstrated the superiority of the approximation in the threshold-based policy. 
Furthermore, we have extended our study to encompass coordinated platooning and adaptive routing in networks.
We decouple the action space by prioritizing routing decisions with travel time estimation.
Subsequently, we employ the aforementioned approximation in the threshold-based policy considering relevant parameters such as travel times to determine speed profiles.
Through simulations in the Nguyen-Dupuis network, our approach has shown better performance compared with conventional methods.
Additionally, we asses the resilience of our approach in dynamically changing networks by simulating edge disconnections, revealing its ability to maintain efficient operations even under such conditions.

This work can be extended in several directions. 
First, incorporating a non-linear function to approximate cost functions would enable the inclusion of additional traffic information, such as density and flow.
Second, the interaction between CAVs and human-driven vehicles need to be analyzed and addressed.
Third, the utilization of traffic flow models, like the cell transmission model (CTM), within the dynamic programming framework holds potential for improving computation efficiency in time-dependent networks.